# First M87 Event Horizon Telescope Results. III.
# Data Processing and Calibration

The Event Horizon Telescope Collaboration
(See the end matter for the full list of authors.)

## Abstract

We present the calibration and reduction of Event Horizon Telescope (EHT) 1.3 mm radio wavelength observations of the supermassive black hole candidate at the center of the radio galaxy M87 and the quasar 3C 279, taken during the 2017 April 5–11 observing campaign. These global very long baseline interferometric observations include for the first time the highly sensitive Atacama Large Millimeter/submillimeter Array (ALMA); reaching an angular resolution of 25 $\mu$as, with characteristic sensitivity limits of $\sim$1 mJy on baselines to ALMA and $\sim$10 mJy on other baselines. The observations present challenges for existing data processing tools, arising from the rapid atmospheric phase fluctuations, wide recording bandwidth, and highly heterogeneous array. In response, we developed three independent pipelines for phase calibration and fringe detection, each tailored to the specific needs of the EHT. The final data products include calibrated total intensity amplitude and phase information. They are validated through a series of quality assurance tests that show consistency across pipelines and set limits on baseline systematic errors of 2% in amplitude and 1° in phase. The M87 data reveal the presence of two nulls in correlated flux density at $\sim$3.4 and $\sim$8.3 G$\lambda$ and temporal evolution in closure quantities, indicating intrinsic variability of compact structure on a timescale of days, or several light-crossing times for a few billion solar-mass black hole. These measurements provide the first opportunity to image horizon-scale structure in M87.

*Key words:* black hole physics – galaxies: individual (M87, 3C279) – galaxies: jets – techniques: high angular resolution – techniques: interferometric

## 1. Introduction

The principle of very long baseline interferometry (VLBI) is to connect distant radio telescopes to create a single virtual telescope. On the ground, VLBI enables baseline lengths comparable to the size of the Earth. This significantly boosts angular resolution, at the expense of having a non-uniform filling of the aperture. In order to reconstruct the brightness distribution of an observed source, VLBI requires cross-correlation between the individual signals recorded independently at each station, brought to a common time reference using local atomic clocks paired with the Global Positioning System (GPS) for coarse synchronization. The resulting complex correlation coefficients need to be calibrated for residual clock and phase errors, and then scaled to physical flux density units using time-dependent and station-specific sensitivity estimates. Once this process is completed, further analysis in the image domain can refine the calibration using model-dependent self-calibration techniques (e.g., Pearson & Readhead 1984; Wilkinson 1989). For more details on the principles of VLBI, see, e.g., Thompson et al. (2017).

At centimeter wavelengths, the technique of VLBI is well established. Correlation and calibration have been optimized over decades, resulting in standard procedures for the processing of data obtained at national and international facility instruments, such as the Very Long Baseline Array[103]

(VLBA), the Australian Long Baseline Array[104] (LBA), the East Asian VLBI Network[105] (EAVN), and the European VLBI Network[106] (EVN). At higher frequencies, the increased effects from atmospheric opacity and turbulence pose major challenges. The characteristic atmospheric coherence timescale is only a few seconds for millimeter wavelengths, and sensitivity must be sufficient to track phase variation over correspondingly short timescales. Large collecting areas and wide recording bandwidths prove essential when observing even the brightest continuum sources over a range of elevations and reasonable weather conditions. Furthermore, the transfer of phase solutions from a bright calibrator to a weak source, typically done at centimeter wavelengths, is not feasible at high frequencies, because differential atmospheric propagation effects are more significant, and because there are few bright, compact calibrators.

The Event Horizon Telescope (EHT) is a global VLBI array of millimeter- and submillimeter-wavelength observatories with the primary goal of studying the strong gravity, near-horizon environments of the supermassive black holes in the Galactic Center, Sagittarius A* (Sgr A*), and at the center of the nearby radio galaxy M87 (Doeleman et al. 2009; EHT Collaboration et al. 2019b, hereafter Paper II). In 2017 April, the EHT conducted science observations at a wavelength of $\lambda \simeq 1.3$ mm, corresponding to a frequency of $\nu \simeq 230$ GHz. The network was joined for the first time by the Atacama Large Millimeter/submillimeter Array (ALMA) configured as a phased array, a capability developed by the ALMA Phasing Project (APP; Doeleman 2010; Fish et al. 2013; Matthews et al. 2018). The addition of ALMA, as a highly sensitive central anchor station, drastically changes the overall characteristics and sensitivity limits of the global array (Paper II).

Although operating as a single instrument spanning the globe, the EHT remains a mixture of new and well-exercised stations, single-dish telescopes, and phased arrays with varying designs and operations. Each observing cycle over the last several years has been accompanied by the introduction of new telescopes to the array, and/or significant changes and upgrades to existing stations, data acquisition hardware, and recorded bandwidth (Paper II). EHT observations result in data spanning a wide range of signal-to-noise ratio (S/N) due to the heterogeneous nature of the array, and the high observing frequency produces data that are particularly sensitive to systematics in the signal chain. These factors, along with the typical challenges associated with VLBI, have motivated the development of specialized processing and calibration techniques.

In this Letter we describe the full data processing pathway and pipeline convergence leading to the first science release (SR1) of the EHT 2017 data. Given the uniqueness of the data set and scientific goal of the EHT observations, our processing focuses on the use of unbiased automated procedures, reproducibility, and extensive review and cross-validation. In particular, data reduction is carried out with three independent phase calibration (fringe-fitting) and reduction pipelines. The Haystack Observatory Processing System (HOPS; Whitney et al. 2004) has been the standard for calibrating EHT data from prior observations (e.g., Doeleman et al. 2008, 2012; Fish et al. 2011, 2016; Akiyama et al. 2015; Johnson et al. 2015; Lu et al. 2018). HOPS reduction of the 2017 data is supported by a suite of auxiliary calibration scripts to form the EHT-HOPS pipeline (Blackburn et al. 2019). The Common Astronomy Software Applications package (CASA; McMullin et al. 2007) is primarily aimed at processing connected-element interferometer data. The recent addition of a fringe fitter and reduction pipeline has enabled the use of CASA for high-frequency VLBI data processing (Janssen et al. 2019a, I. van Bemmel et al. 2019, in preparation). The NRAO Astronomical Image Processing System (AIPS; Greisen 2003) is the most commonly used reduction package for centimeter VLBI data. For this work, an automated ParselTongue (Kettenis et al. 2006) pipeline was constructed and tailored to the needs of EHT data reduction in AIPS.

The SR1 data consist of Stokes I complex interferometric visibilities of M87 and the quasar 3C 279, corresponding to spatial frequencies of the sky brightness distribution sampled by the interferometer. M87 data indicate the presence of a resolved compact emission structure on a spatial scale of a few tens of $\mu$as, persistent throughout the week-long observing campaign. Closure phases and closure amplitudes unambiguously reflect non-trivial brightness distributions on M87 for the first time. They display broad consistency over different days, and in certain cases show clear evolution. A detailed analysis of this near-horizon-scale structure is the subject of companion Letters (EHT Collaboration et al. 2019a, 2019c, 2019d, 2019e, hereafter Papers I, IV, V, and VI, respectively).

This Letter is organized as follows. Section 2 presents an overview of the 2017 April observations. In Section 3 we outline the data flow from observations to science-ready data sets. We describe the correlation process in Section 4, the phase calibration process via three independent fringe-fitting pipelines in Section 5, and the common flux density calibration scheme and amplitude error budget in Section 6. We give an overview of SR1 data products and a rudimentary description of their most evident, remarkable properties in Section 7. We present data set validation procedures and tests, estimates of systematic errors, and inter-pipeline comparisons in Section 8. Conclusions are given in Section 9.

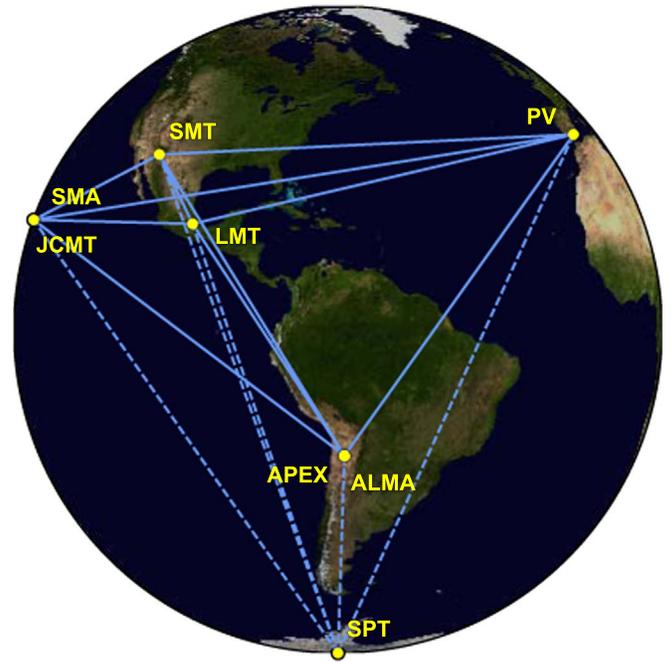

**Figure 1.** The eight EHT 2017 stations over six geographic locations as viewed from the equatorial plane. Solid baselines represent mutual visibility on M87 ($+12°$ decl.), while dashed baselines to SPT are also present for 3C 279 ($-6°$ decl.).

## 2. Observations

The EHT 2017 science observing run was scheduled for 5 nights during the 10-night 2017 April 5–14 (UTC) window with eight participating observatories at six distinct geographical locations, shown in Figure 1: the ALMA and the Atacama Pathfinder Experiment (APEX) in the Atacama Desert in Chile, the Large Millimeter Telescope Alfonso Serrano (LMT) on the Volcán Sierra Negra in Mexico, the South Pole Telescope (SPT) at the geographic south pole, the IRAM 30 m telescope (PV) on Pico Veleta in Spain, the Submillimeter Telescope (SMT) on Mt. Graham in Arizona, and the Submillimeter Array (SMA) and the James Clerk Maxwell Telescope (JCMT) on Maunakea in Hawai'i. A detailed description of the EHT array is presented in Paper II. The 2017 science observing run consisted of observations of six science targets: the primary EHT targets Sgr A* and M87, and the secondary targets 3C 279, OJ 287, Centaurus A, and NGC 1052.

An array-wide go/no-go decision was made a few hours before the start of each night's schedule, based on weather conditions and technical readiness at each of the participating observatories. A dry run of the go/no-go decision making was performed on April 4 to assess triggering and readiness procedures. All sites were technically ready and with good weather on the first night of the observing window. Observations were triggered on 2017 April 5, 6, 7, 10, and 11. Table 1 shows the median zenith sky opacities for each of the triggered days. April 8 was not triggered due to thunderstorms at the LMT, SMT shutdown due to strong winds, and the need to run

| Station | Median Zenith $\tau_{1.3\ mm}$ | | | | |
| | Apr 5 | Apr 6 | Apr 7 | Apr 10 | Apr 11 |
| --- | --- | --- | --- | --- | --- |
| ALMA/APEX | 0.06 | 0.04 | 0.05 | 0.03 | 0.06 |
| SMA/JCMT | 0.10 | 0.07 | 0.09 | 0.05 | 0.08 |
| PV | 0.18 | 0.13 | 0.14 | 0.10 | 0.15 |
| LMT | 0.13 | 0.16 | 0.21 | 0.26 | 0.24 |
| SMT | 0.21 | 0.28 | 0.23 | 0.19 | 0.16 |
| SPT | 0.04 | 0.05 | 0.07 | 0.08 | 0.07 |

**Note.** Median zenith sky opacities are measured at each site and reported through station log files and the VLBImonitor as described in Paper II.

technical tests at ALMA. April 9 was not triggered due to a chance of the SMT remaining closed due to strong winds and LMT snow forecast. Weather was good to excellent for all other stations throughout the observing window.

In addition to favorable weather conditions, operations at all sites were successful and resulted in fringe detections across the entire array. A number of mild to moderate site and data issues were uncovered during the analysis, and their detailed characterization and mitigation are given in the Appendix. Notable issues affecting processing, calibration, and data interpretation are: (1) a clock frequency instability at PV resulting in ~50% amplitude loss to that station; (2) recorder configuration issues at APEX resulting in a significant number of data gaps and low data validity at correlation; (3) pointing errors at LMT, large compared to the beam, resulting in unpredictable amplitude loss and inter- and intra-scan gain variability; and (4) a common local oscillator (LO) used at SMA and JCMT resulting in opposite sideband contamination at the level of ~15% for short integration times, making the SMA–JCMT intra-site baseline less useful for calibration. All known issues with a significant effect on the data are addressed at various stages of processing and calibration, although some (such as residual gains at the LMT, and SMA–JCMT sideband contamination) necessitate additional care taken during data interpretation.

M87 ($\alpha_{J2000} = 12^h30^m49\overset{s}{.}42$, $\delta_{J2000} = 12°23'28''04$) was observed as a target source on three nights (2017 April 5, 6, and 11). In addition, seven scans on M87 were included as a calibration source (for 3C 279) on 2017 April 10. Each of the four tracks consists of multiple scans lasting between 3 and 7 minutes. In most tracks, VLBI scans on M87 began when it rose at the LMT and ended when it set below 20° elevation at ALMA. Scans on M87 were interleaved with scans on the quasar 3C 279 ($\alpha_{J2000} = 12^h56^m11\overset{s}{.}17$, $\delta_{J2000} = -05°47'21''52$), another EHT target with a similar R.A. The observed schedules for M87 and 3C 279 during the 2017 campaign are shown in Figure 2. The schedules were optimized for wide $(u, v)$ coverage on all target sources when possible. All stations apart from the JCMT observed with full polarization. The JCMT observed a single circular polarization component per night (right circular polarization (RCP) for April 5 and 6, left circular polarization (LCP) for April 10 and 11).

The 2017 observing run recorded two 2 GHz bands, low and high, centered at sky frequencies of 227.1 and 229.1 GHz, respectively, onto Mark 6 VLBI recorders (Whitney et al. 2013) at an aggregate recording rate of 32 Gbps with 2-bit sampling. All telescopes apart from ALMA observed in circular polarization with the installation of quarter-wave plates. Single-dish sites used block downconverters to convert the intermediate frequency (IF) signal from the front-ends to a common 0–2 GHz baseband, which was digitally sampled via Reconfigurable Open Architecture Computing Hardware 2 (ROACH2) digital backends (R2DBEs; Vertatschitsch et al. 2015). The SMA observed as a phased array of six or seven antennas, for which the phased-sum signal was processed in the SMA Wideband Astronomical ROACH2 Machine (SWARM) correlator (see Primiani et al. 2016; Young et al. 2016, for more details). ALMA observed as a phased array of usually 37 dual linear polarization antennas, for which the phased-sum signal was processed in the Phasing Interface Cards installed at the ALMA baseline correlator (see Matthews et al. 2018 for more details). Instrumentation development leading up to the 2017 observations is presented in Paper II.

## 3. Data Flow

The EHT data flow from recording to analysis is outlined in Figure 3. Through the receiver and backend electronics at each telescope, the sky signal is mixed to baseband, digitized, and recorded directly to hard disk, resulting in petabytes of raw VLBI voltage signal data. The correlator uses an a priori Earth geometry and clock/delay model to align the signals from each telescope to a common time reference, and estimates the pair-wise complex correlation coefficient ($r_{ij}$) between antennas. For signals $x_i$ and $x_j$ between stations $i$ and $j$

$$r_{ij} = \frac{\langle x_i x_j^* \rangle}{\eta_Q \sqrt{\langle x_i x_i^* \rangle \langle x_j x_j^* \rangle}}, \tag{1}$$

where $\eta_Q$ represents a digital correction factor to compensate for the effects of low-bit quantization. For optimal 2-bit quantization, $\eta_Q \approx 0.88$.

The correlation coefficient may vary with both time and frequency. For FX correlators, signals from each antenna are first taken to the frequency domain using temporal Fourier transforms on short segments (F), and then pair-wise correlated (X). The expectation values in Equation (1) are calculated by averaging over time–frequency volumes where the inner products remain stable. At millimeter wavelengths, a correlator can average around 1 s × 1 MHz, or 2 × 10⁶ samples, before clock errors such as residual delay, delay-rate (e.g., Doppler shift), and stochastic changes in atmospheric path length cause unwanted decoherence in the signal (Section 4). The post-correlation data reduction pipeline models and fits these residual clock systematics, allowing data to be further averaged by a factor of 10³ or more, to the limits imposed by intrinsic source structure and variability (Section 5). For many EHT baselines, the astronomical signal is not detectable above the noise until phase corrections resulting from these calibration solutions are applied and the data are coherently (vector) averaged.

In addition to reducing the overall volume and complexity of the data, the calibration process attempts to relate the pair-wise correlation coefficients $r_{ij}$, which are in units of thermal noise of the detector, to correlated flux density in units of Jansky (Jy),

$$r_{ij} = \gamma_i \gamma_j^* \ V_{ij}. \tag{2}$$

The *visibility function*, $V_{ij}$, represents the mutual coherence of the electric field between ends of the baseline vector joining the sites, projected onto the plane of propagation. For an ideal interferometer, $V_{ij}$ samples a Fourier component of the brightness distribution on the sky (via the van Cittert–Zernike theorem; van Cittert 1934; Zernike 1938; Thompson et al. 2017). The dimensionless spatial frequency $\boldsymbol{u} = (u, v)$ of the Fourier component is determined by the projected baseline expressed in units of the observing wavelength. Here, we have made the implicit assumption that the relationship between correlation coefficient and visibility can be factored into complex station-based forward gains $\gamma_i$ and $\gamma_j$. This process of flux density calibration requires an a priori assessment of the sensitivity of each antenna in the array, captured by the *system-equivalent flux density* (SEFD$_i = |1/\gamma_i|^2$) of the thermal noise power, as described in Section 6.

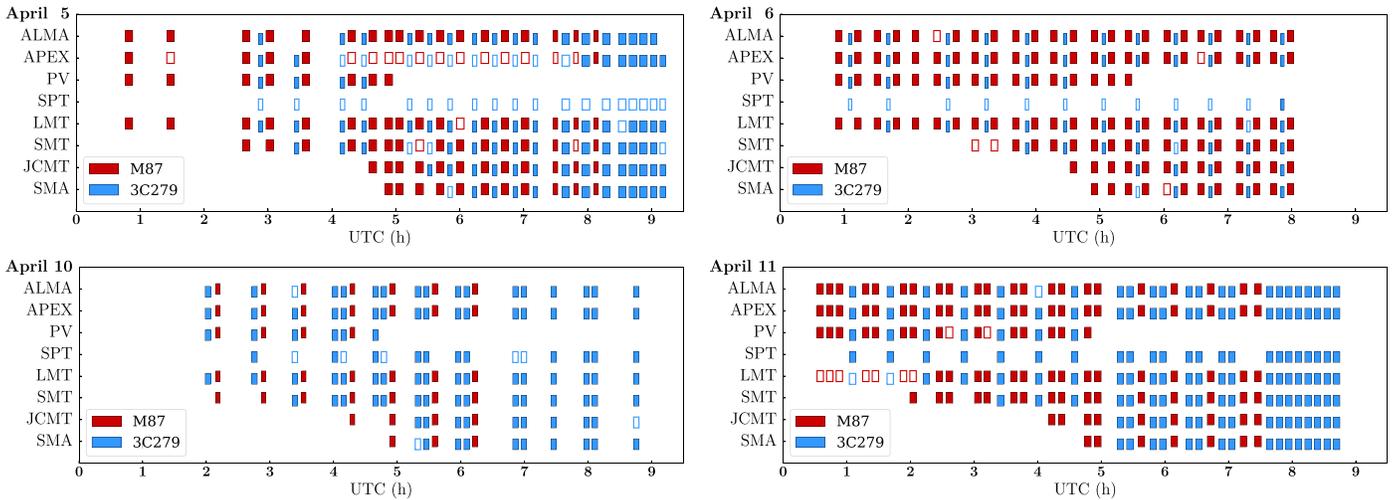

**Figure 2.** EHT 2017 observing schedules for M87 and 3C 279 covering the four days of observations. Empty rectangles represent scans that were scheduled, but were not observed successfully due to weather, insufficient sensitivity, or technical issues. The filled rectangles represent scans corresponding to detections available in the final data set. Scan duration varies between 3 and 7 minutes, as reflected by the width of each rectangle.

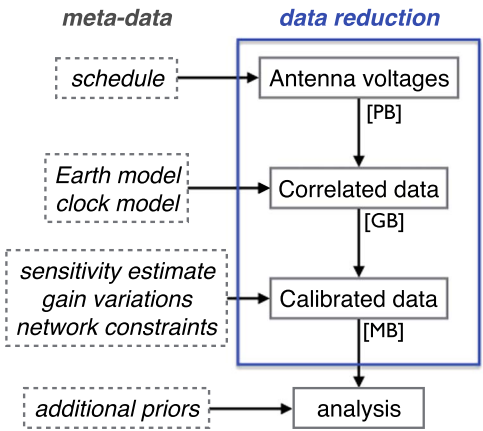

**Figure 3.** Data processing pathway of an EHT observation from recording to source parameter estimation (images, or other physical parameters). At the calibration stage, instrumental and environmental gain systematics are estimated and removed from the data so that a smaller and simpler data product can be used for source model fitting at a downstream analysis stage.

After the basic calibration and reduction process, the data are passed through additional post-processing tasks to further average the data to a manageable size for source imaging and model fitting, and to apply any network self-calibration constraints based on independent a priori assumptions about the source, such as large-scale (milliarcsecond and larger) structure, total flux density, and degree of total polarization

(Section 6.2). The final network-calibrated data products are further averaged to a 10 s segmentation in time and across each 2 GHz band to provide smaller files for downstream analysis (Section 7.1).

## 4. Correlation

The recorded data from each station were split by frequency band and sent to MIT Haystack Observatory and the Max-Planck-Institut für Radioastronomie (MPIfR) for correlation, as described in Paper II. The Haystack correlator handled the low-frequency band (centered at 227.1 GHz), with MPIfR correlating the high band (centered at 229.1 GHz). Each correlator is a networked computer cluster running a standard installation of the `DiFX` software package (Deller et al. 2011). The correlators use a model (`calc11`) of the expected wavefront arrival delay as a function of time on each baseline. The delay model very precisely takes into account the geometry of the observing array at the time of observation, the direction of the source, and a model of atmospheric delay contributions (e.g., Romney 1995). Baseband data on a few high-S/N scans with good coverage were exchanged between the two sites to verify the output of each correlator against the other.

Data were correlated with an accumulation period (AP) of 0.4 s and a frequency resolution of 0.5 MHz (Figure 4). Due to the need to rationalize frequency channelization within the ALMA setup (each 1.875 GHz spectral window at ALMA is broken up into 32 spectral IFs of 62.5 MHz, separated by 58.59375 MHz and thus slightly overlapping; Matthews et al. 2018), the frequency points are grouped into IFs that are 58 MHz wide (using `DiFX` *zoom mode*), each with 116 individual channels and a small amount of bandwidth discarded between spectral IFs.

At the SMA, the original data are recorded in the frequency domain rather than the time domain, owing to the architecture of the SMA correlator. Moreover, the recorded frequency range of 2288 MHz is slightly larger and offset by 150 MHz from the frequency range at the other non-ALMA sites. An offline preprocessing pipeline, called the Adaptive Phased-array and Heterogeneous Interpolating Downsampler for SWARM (APHIDS; Primiani et al. 2016), is used to perform the necessary filtering, frequency conversion, and transformation to the time domain, so that the format of the SMA data delivered to the VLBI correlator is the same as for single-dish stations. Part of the necessary offline pre-processing includes deriving clock offsets on a scan-by-scan basis for the delivered data. These offsets are determined by cross-correlating the pre-processed SMA data with separate data recorded with an R2DBE-Mark 6 pair, taking a second IF signal from the SMA reference antenna as input.

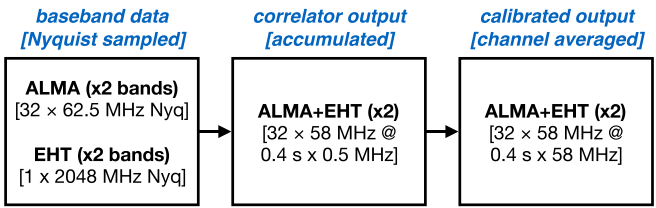

**Figure 4.** Time and frequency resolution of EHT 2017 data as it is recorded and processed. Correlation parameters for the EHT are chosen to be compatible with ALMA's recorded sub-bands that are 62.5 MHz wide, overlap slightly, and have starting frequencies aligned to $1/(32\,\mu s)$. The raw output after calibration and reduction maintains the original correlator accumulation of 0.4 s, but averages over each 58 MHz spectral IF, centered on each ALMA sub-band. The data are further averaged at the network amplitude self-calibration stage (not shown) for a more manageable data volume.

The IF from the JCMT was recorded using backend equipment installed at the SMA (Paper II). This was achieved by transporting the first IF from the JCMT to the SMA, where the second downconversion, digitization, and recording were done. Because the second downconversion at the SMA introduces a net offset of 150 MHz with respect to the nominal EHT RF band, this means that the recorded JCMT data sent to the correlator are subject to the same frequency offset. The mismatch eliminates one of the thirty-two 58 MHz spectral IFs in the final correlation for JCMT baselines.

ALMA observes linear polarization, while the rest of the EHT observes circular polarization. The software routine `PolConvert` (Martí-Vidal et al. 2016; Matthews et al. 2018) was created to convert visibilities, output from the correlator in a mixed-polarization basis, to the pure circular basis of the EHT. `PolConvert` takes auxiliary calibration input from the quality assurance stage 2 (QA2) ALMA interferometric reduction of data (Goddi et al. 2019). Execution of the `PolConvert` tool completes the correlation (circularized visibilities on baselines to ALMA) and provides final ANTAB[107] format data for flux density calibration of the ALMA phased array. The original native (Swinburne format) correlator output from `DiFX` is converted into available `DiFX` tools to a `Mark4` (Whitney et al. 2004) compatible file format for processing through HOPS, and to `FITS-IDI` (Greisen 2011) files for further processing with AIPS and CASA.

## 5. Fringe Detection

In the limit for which all correlator delay model parameters were known perfectly ahead of time and there were no atmospheric variations, the model delays would exactly compensate for the delay on each baseline of the data, and the correlated data could be coherently integrated in time and frequency to build up sensitivity. In practice, many of the model parameters are not known exactly at correlation. For example, the observed source may have structure and may be centered at an offset from the expected coordinates, the position of each telescope may differ from the best estimate, instrumental electronic delays may not be known, or variable water content in the atmosphere may cause the atmospheric delay to deviate from the simple model. It is therefore necessary to search in delay and delay-rate space for small corrections to the model values that maximize the fringe amplitude: in VLBI data processing this process is known as fringe-fitting (e.g., Cotton 1995). In this section, we describe three independent fringe-fitting pipelines for phase calibration, based on three different software packages for VLBI data processing: HOPS (Section 5.1), CASA (Section 5.2), and AIPS (Section 5.3).

### 5.1. HOPS Pipeline

HOPS[108] is a collection of software packages and data framework designed to analyze and reduce output from a Mark III, IV, or `DiFX` correlator. It has been used extensively for the processing of early EHT data (Doeleman et al. 2008, 2012; Fish et al. 2011, 2016; Akiyama et al. 2015; Johnson et al. 2015; Lu et al. 2018). For EHT 2017 observations, HOPS was augmented with a collection of auxiliary calibration scripts, and packaged into an EHT-HOPS pipeline (Blackburn et al. 2019) for automated processing of this and similar data sets. Compared to the reduction of data from previous runs, the EHT-HOPS pipeline is unique in that it finds a single self-consistent global fringe solution (station-based delays, delay-rates, and instrumental and atmospheric phase) for calibration. The pipeline also provides standard `UVFITS` formatted visibility data products for downstream analysis.

The EHT-HOPS pipeline processes output from the `DiFX` correlator that has been converted to `Mark4` format via the `DiFX` tool `difx2mark4`. This conversion process includes normalization by auto-correlation power per 58 MHz spectral IF in each AP of 0.4 s (Figure 4), as well as a 1/0.88252 amplitude correction factor for 2-bit quantization efficiency. Stages of the pipeline (Figure 5) run the HOPS fringe fitter `fourfit` several times (once per stage) while making iterative corrections to the phase calibration applied to the data before solving for delays and delay-rates. The initial setup (*default config, flags*—Figure 5) includes manual flagging (removal of bad data) in time and frequency, as well as an ALMA-specific correction for digital phase offsets between spectral IFs.

ALMA is used as a reference station for estimating stable instrumental phase (*phase bandpass*) and relative delay between right and left circular polarization (*R-L delay offsets*) for remote stations. The estimates are done using S/N-weighted averages of the strong ALMA baseline measurements. Here we make use of the fact that ALMA RCP and LCP data are already delay- and phase-calibrated during the QA2/ `PolConvert` process (Goddi et al. 2019). For rapid nonlinear phase (*atmospheric phase*) that varies over seconds and that must be calibrated on-source, the strongest station (generally ALMA when it is present; see also Section 2 of Paper II) is automatically determined for each scan based on signal-to-noise, and is used as a phase reference. Baselines to the reference station are then used to phase stabilize the remaining sites.

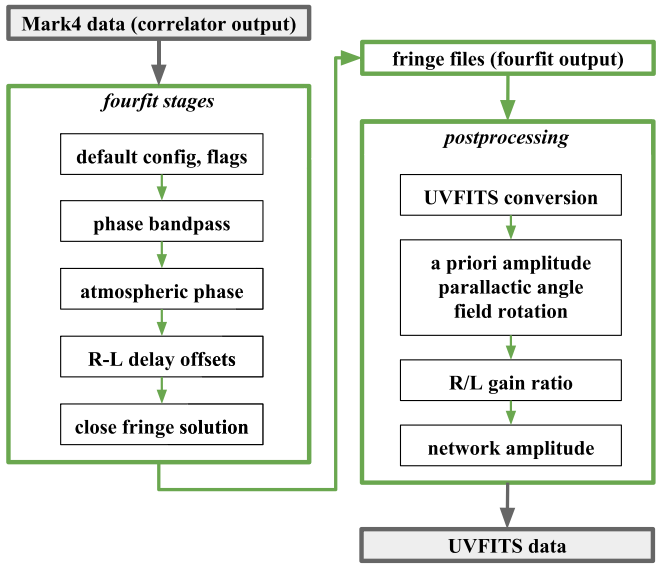

**Figure 5.** Stages of the EHT-HOPS pipeline and post-processing steps, as described in the text. The first five stages, shown in the left box, are iterations of HOPS fringe fitter `fourfit`. Here, a comprehensive phase calibration model is gradually built for the data. At the end of the five fourfit stages, the correlation coefficients are evaluated at a single global (station-based) set of relative delays and delay-rates. The data are then converted to `UVFITS` format, and a remaining suite of post-processing tools provide amplitude calibration and time-and-polarization-dependent phase calibration.

Due to the large number of free parameters involved in correcting for atmospheric phase, a leave-one-out cross-estimation approach is adopted for this step to avoid self-tuning. For each baseline, a smooth phase model is estimated by stacking RCP and LCP data over 31 (of 32) spectral IFs. The estimated phase from the 31-IF average is used to correct the remaining IF, and the process cycles through IFs to cover the full band. In this way, phase corrections are never estimated from the same data to which they are applied, which avoids introducing false coherence from self-tuning to random thermal noise and introducing a positive bias to amplitudes. The effective solution interval for the phase model depends on S/N, and is chosen per baseline to balance anticipated atmospheric phase drift with thermal noise in the estimate. Additional a priori corrections for small residual clock frequency offsets after correlation (Appendix) are made here as well.

During a final reduction with `fourfit` (close fringe solution), rather than fitting for unconstrained delays and delay-rates per baseline and polarization product, a single set of station-based delays and delay-rates is fixed corresponding to a global fringe solution. These are derived from a least-squares solution (as proposed by Alef & Porcas 1986) to relative delays and delay-rates from confident baseline detections with S/N > 7, and stations that remain unconstrained by this process are removed from the data set. No interpolation of these fringe solutions is performed across sources and scans; instead, precise closure of delay and delay-rate from strong baseline detections is required to report any measurement on a weak baseline. Correlation coefficients on baselines with no detectable signal are still calculated (Figure 11, where S/N < few), but only when the relative clock model is constrained through other baseline detections.

The resulting complex visibility data are converted to `UVFITS` format, and amplitude calibration is done in the

EHT Analysis Toolkit's (`eat`)[109] post-processing framework, shared by all pipelines and described in Section 6. For the HOPS pipeline, the calibration of complex polarization gain ratios is performed in a post-processing stage rather than during `fourfit`. Deterministic field rotation from parallactic angle and receiver mount type is corrected as a complex polarization-dependent a priori gain factor, and a smoothly varying polynomial model is fit over many sources and used to correct residual RCP−LCP phase drift for each station. Details for all steps can be found in Blackburn et al. (2019).

The EHT-HOPS pipeline was additionally used for the reduction of observations of Sgr A* and calibrators at 86 GHz, with the Global Millimeter VLBI Array[110] (GMVA) joined by ALMA. Despite the magnitude difference in bandwidth, a similar reduction to EHT data was performed on the GMVA data set. ALMA baselines were used to estimate stable instrumental phase and delay corrections. Baselines to either ALMA or the Green Bank Telescope (GBT) were used, due to their high S/N, to correct for stochastic atmospheric phase fluctuations on timescales of a few seconds. The performance of the pipeline on the GMVA data is described in Blackburn et al. (2019) while scientific results from the data set are validated against historical observations in Issaoun et al. (2019).

### 5.2. CASA Pipeline

The `CASA` (McMullin et al. 2007) package was developed by NRAO to process data acquired with the JVLA and ALMA connected-element interferometers and in recent years has become the standard software for the calibration and analysis of radio-interferometric data. A newly developed fringe-fitting task `fringefit` (I. van Bemmel et al. 2019, in preparation) has added the necessary delay and delay-rate calibration capabilities for VLBI. The modular, general-purpose `rPICARD` VLBI data reduction pipeline (Janssen et al. 2019a) is used for the calibration of EHT data. This section describes the incremental `rPICARD` calibration steps for EHT data, summarized in Figure 6.

The `importfitsidi` CASA task is used to import the `FITS-IDI` correlator output into CASA. Additionally, a digital correction factor for the 2-bit recorder sampling is applied when the data are loaded. Bad data are flagged based on text files compiled from station logs and known sources of radio frequency interference in stations' signal chains with the `flagdata` task before performing the incremental calibration procedures. The `accor` task is used to scale the auto-correlations to unity and adjust the cross-correlations accordingly, correcting for incorrect sampler settings from the data recording stage. This is done for each 58 MHz spectral IF individually, thereby correcting for a coarse bandpass at each station. This amplitude bandpass is refined by dividing the data by the auto-correlations at the 0.5 MHz channel resolution.

The phase calibration is done with the `fringefit` task, which solves for station-based residual post-correlation phases, delays, and rates with respect to a chosen reference station (Schwab & Cotton 1983). Unlike the HOPS pipeline, where field rotation angles are corrected a posteriori, `rPICARD` applies field rotation angle gain solutions on-the-fly, i.e., before each phase calibration correction. The most sensitive station is picked as reference in each scan. Eventually, all fringe solutions are re-referenced with the CASA `rerefant` task to a common station for each observing track to ensure phase continuity across scans.

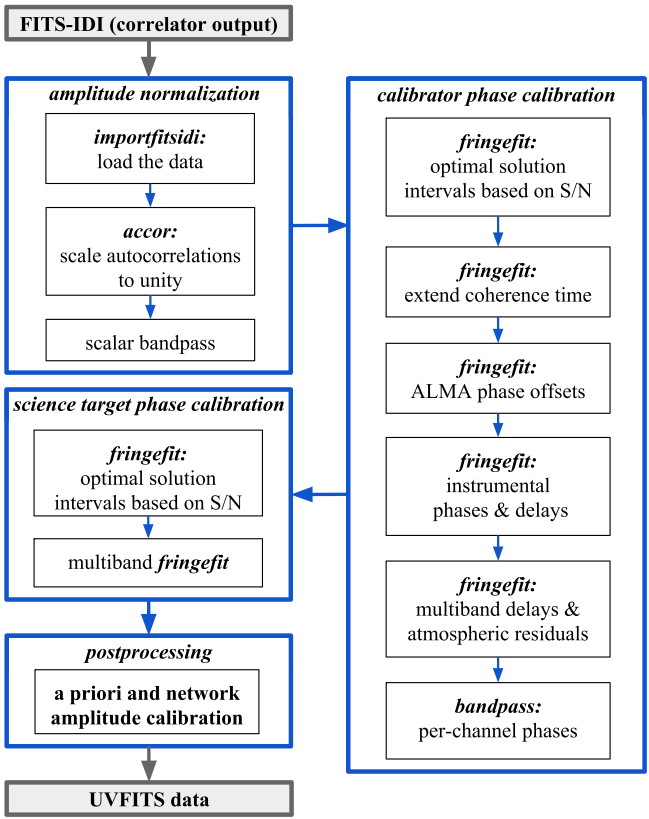

Phases are first calibrated for the high S/N calibrator sources, which are used to correct for instrumental effects. Optimal time solution intervals to calibrate atmospheric intra-scan phase fluctuations ($\mathcal{T}_{sol}$) are determined automatically based on the S/N of the data. The search is done for short solution intervals, close to the coherence time, which still yield detections on all possible baselines (Janssen et al. 2019a). Typical solution intervals range from 2 to 30 s. Using these solution intervals, phases and rates are calibrated to extend the coherence time of the calibrator scans. This results in high S/N scan-based fringe solutions per 58 MHz spectral IF, which are used to obtain calibration solutions for instrumental effects. ALMA-induced phase offsets between spectral IFs are corrected with the short ALMA–APEX baseline. All baselines in the array are used by the global fringe fitter in the next step to solve for residual instrumental phase and delay offsets for all stations. After removing these instrumental data corruptions, a final `fringefit` step solves for multi-band delays on the (previously determined) solution intervals. A 60 s median window filter is used to smooth the slowly varying multi-band delays, which effectively removes potential outliers. After fringe fitting, the phases are coherent in time and frequency, and the `bandpass` task is used to solve for the frequency-dependent phase gains within each 58 MHz spectral IF for each station, using the combined data of all calibrator sources.

After all instrumental effects are calibrated out, the optimal fringe-fit solution intervals $\mathcal{T}_{sol}$ are determined for the weaker science targets, and phases, delays, and rates are solved for in a single `fringefit` step. The intra-scan fringe fritting on short solution intervals flags low S/N segments where no fringes are found to a specific station, e.g., when a station arrived late on source. Finally, the `exportuvfits` task is used to export the calibrated data from internal *Measurement Set* format to `UVFITS` files, which are then flux-density and network-calibrated in the common post-processing framework.

Janssen et al. (2019a) demonstrate the `rPICARD` calibration capabilities in a close comparison with a traditional AIPS-based calibration using 43 GHz VLBA data of M87. The resultant image of the jet and counter-jet, which reveals a complex collimation profile, is in good agreement with earlier results from the literature (e.g., Walker et al. 2018). The `rPICARD` pipeline was further used for the generation of synthetic EHT data (Paper IV), where known input delay and phase offsets were recovered as a ground-truth validation.

### 5.3. AIPS Pipeline

AIPS (Greisen 2003) is the most widely used software package for VLBI data reduction and processing at frequencies at or below ∼86 GHz. It is commonly used in the VLBI community and was built to process low-S/N data from fairly homogeneous centimeter-wave observatories at low recording bandwidths. The EHT, however, falls in a different category: its high recording bandwidth and heterogeneous array produce data with a wide range of S/N, often dominated by systematic effects instead of thermal noise. These properties required the development of a custom pipeline based on AIPS, deviating from standard fringe-fitting procedures for lower frequency data processing as outlined in e.g., the AIPS Cookbook.[111]

The custom AIPS pipeline is an automated `Python`-based script using functions implemented in the `eat` package. It makes use of `ParselTongue` (Kettenis et al. 2006), which provides a platform to manipulate AIPS tasks and data outside of the AIPS interface. The pipeline is summarized in Figure 7 and shows individual tasks used for calibration. A suite of diagnostic plots, using tasks `VPLOT` and `POSSM`, are also generated at each calibration step within the pipeline.

The loading of EHT data into AIPS, during which digital corrections for 2-bit quantization efficiency are applied, requires a concatenation of several packaged `FITS-IDI` files and a careful handling of the JCMT, which observes with a slightly shifted IF setup of the band (Section 4). The pipeline reduces each band (low and high) in separate runs. Data inspection and flagging of spurs in the frequency domain from accumulated scalar bandpass tables (generated with `BPASS`) and dropouts or amplitude jumps in the time domain are done interactively with the AIPS tasks `BPEDT` and `EDITA`. The flags are saved in output flag tables to use in non-interactive reruns of the pipeline. Standard amplitude normalization steps are performed with the AIPS task `ACSCL`. The field rotation angle corrections are performed with an EHT-specific receiver mount correction script (`ehtutil.ehtpang`, modifying the antenna table from the `DiFX` alt-az default to the proper receiver mounts of each station) using the AIPS task `CLCOR` before fringe fitting.

---

[111] http://www.aips.nrao.edu/cook.html

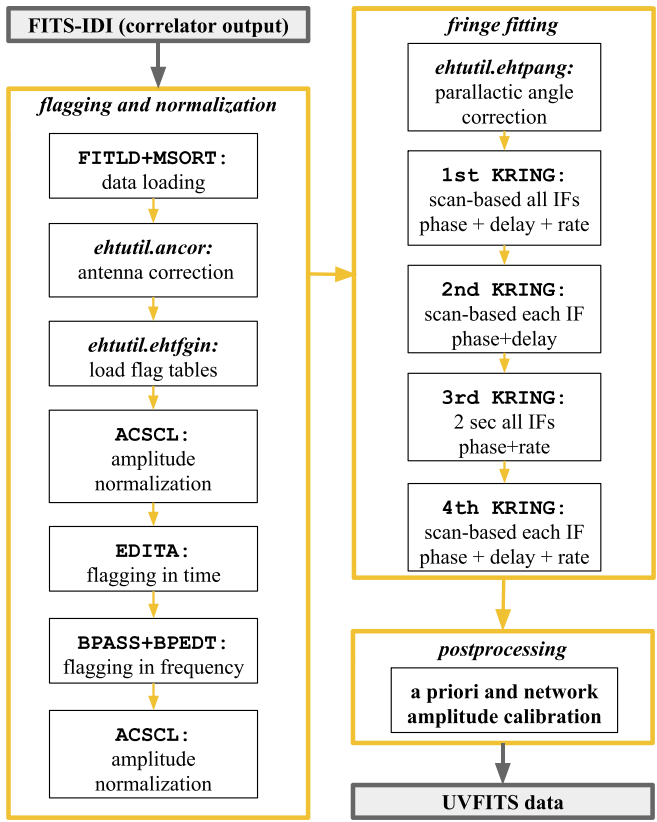

**Figure 7.** Stages of the AIPS fringe-fitting pipeline and post-processing steps. The pipeline begins with direct data editing (interactively or via input correction and flag tables) and amplitude normalization (first box). The phase calibration process then follows via four steps with the AIPS fringe fitter `KRING` to solve for phase and delay offsets and rates (second box). Finally, post-processing steps are done outside of AIPS for amplitude calibration (third box).

The fringe-fitting steps follow a similar framework to the HOPS pipeline but use `KRING`,[112] a station-based fringe fitter that outperforms the standard `FRING` in terms of computational efficiency for large data sets, while maintaining an equivalent accuracy. The first step of the fringe search, commonly known as instrumental phase calibration, consists of solving for delay and phase offsets and fringe-rates using the full scan coherence and full 2 GHz bandwidth (combining spectral IFs). The second step solves for delay and phase offset residuals per individual spectral IF, again using the full scan coherence. The third step uses a fixed solution interval of 2 s to solve for fast phase rotations in time across the full bandwidth (combining IFs). The final stage is solving for scan-based residual delays and phases per individual spectral IF.

The AIPS pipeline particularly relies on ALMA being present to accurately solve for short interval solutions, as it uses ALMA as the reference station for the initial baseline-based FFT within `KRING`. Without ALMA, or in certain cases of a weak baseline to ALMA, `KRING` is unable to accumulate enough S/N in a single spectral IF or within a two-second segment to constrain a fringe solution. After applying all calibration steps, the data are frequency-averaged and exported in `UVFITS` format. A priori and network calibration are performed outside of AIPS in the common post-processing framework.

## 6. Flux Density Calibration

The flux density calibration for the EHT is done in two steps and is a common post-processing procedure for all three phase calibration pipelines, as it involves very little handling of the data themselves. In Section 6.1, we describe the a priori calibration process to calibrate visibility amplitudes to a common flux density scale across the array. In Section 6.2, we present the network calibration process, where we use array redundancy to absolutely calibrate stations with an intra-site companion.

### 6.1. A Priori Amplitude Calibration

A priori amplitude calibration serves to calibrate visibility amplitudes from correlation coefficients to flux density measurements, as in Equation (2). As the normalized correlation coefficients are in units of noise power, it is necessary to account for telescope sensitivities to convert to a uniform flux density scale across the array. The SEFD of a radio telescope is the total system noise represented in units of equivalent incident flux density above the atmosphere. It can be written as

$$\mathrm{SEFD} = \frac{T_{\mathrm{sys}}^*}{\mathrm{DPFU} \times \eta_{\mathrm{el}}}, \tag{3}$$

using the three measurable parameters:

1. $T_{\mathrm{sys}}^*$: the *effective system noise temperature* describes the total noise characterization of the system corrected for atmospheric attenuation (Equations (4) and (5)),
2. DPFU: the *degrees per flux density unit* provides the conversion factor (K/Jy) from a temperature scale to a flux density scale, correcting for the aperture efficiency (Equation (6)),
3. $\eta_{\mathrm{el}}$: the *gain curve* is a modeled elevation dependence of the telescope's aperture efficiency (Equation (7)), factored out of the DPFU to track gain variation as the telescope moves across the sky.

The EHT is a heterogeneous array with telescopes of various sensitivities (ranging nearly three orders of magnitude, see Figure 8), operation schemes, and designs. A clear understanding of each station's metadata measurement and delivery is required for an accurate calibration of the measured visibilities. We determine the SEFDs of the individual stations and their uncertainties under idealized conditions, assuming adequate pointing and focus (see Sections 6.1.1, 6.1.3, and 6.1.4). Further losses and uncertainty in the SEFDs, particularly those induced by focus or pointing errors, are difficult to quantify using available metadata, but are qualitatively explained in Section 6.1.5. A more quantitative assessment of station behavior can be done via derived residual station gains from self-calibration methods in imaging or model fitting (Papers IV, VI).

### 6.1.1. Quantifying Station Performance

In order to determine the sensitivity of a single-dish station at a given time, measurements of the effective system temperature, the DPFU, and the gain curve are required. Here we provide details on how these parameters are measured for the EHT array.

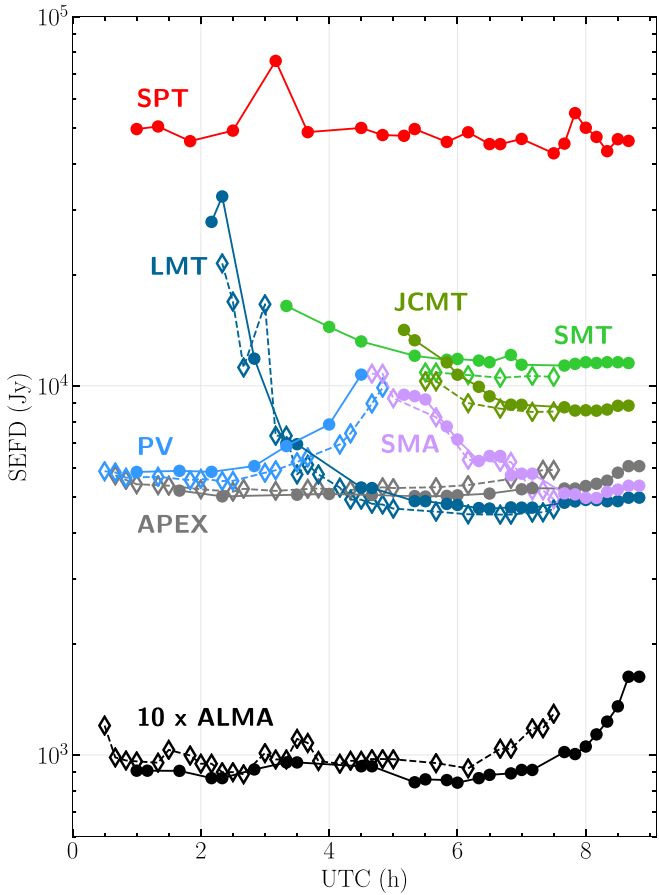

**Figure 8.** Example of SEFD values during a single night of the 2017 EHT observations (April 11, low-band RCP). Values for 3C 279 are marked with full circles, values for M87 are marked with empty diamonds. ALMA SEFDs have been multiplied by 10 in this plot. The SPT is observing 3C 279 at an elevation of just 5°.8, resulting in an uncharacteristically high SEFD due to the large airmass.

The EHT operates in the millimeter-wave radio regime, where observations are very sensitive to atmospheric absorption and water vapor content. In contrast with centimeter-wave interferometers (e.g., VLBA/JVLA), millimeter-wave telescopes typically measure $T_{sys}^*$ via the "chopper" (or hot-load) method: an ambient temperature load $T_{hot}$ with known blackbody properties is placed in front of the receiver, blocking everything but the receiver noise, and the resulting noise power is compared to the same measurement on cold sky. Assuming $T_{hot} \sim T_{atm}$ (the hot load is at a temperature comparable to the radiating atmosphere), this method automatically compensates for atmospheric absorption to first order, essentially transferring the incident flux density reference point to above the atmosphere (e.g., Penzias & Burrus 1973; Ulich & Haas 1976):

$$T_{sys}^* \simeq e^\tau (T_{rx} + (1 - e^{-\tau})T_{atm}), \tag{4}$$

where $T_{rx}$ is the receiver noise temperature, and $\tau$ is the sky opacity in the line of sight. Details on the chopper techniques adopted for the EHT are provided in a technical memo[113] (Issaoun et al. 2017a).

Three stations in the EHT array have double-sideband (DSB) receivers in 2017 (SMA, JCMT, and LMT), where both upper and lower sidebands on either side of the oscillator frequency are folded together in the recorded signal (e.g., Iguchi 2005, Paper II). Because only one 4 GHz sideband is correlated across the array, we correct $T_{sys}^*$ for the excess noise contribution from the uncorrelated sideband

$$T_{sys}^* = T_{sys,DSB}^*(1 + r_{sb}), \tag{5}$$

where the sideband ratio $r_{sb}$ is the ratio of source signal power in the uncorrelated sideband to that in the correlated sideband. A sideband ratio of unity, for an ideal DSB system, is assumed for the SMA and LMT based on known receiver performance. A measured sideband ratio of 1.25 is used for the JCMT.[114] The remaining stations use sideband-separating receiver systems and do not need this adjustment. The SPT, although sideband-separating, is believed to have suffered from a degree of incomplete sideband separation in 2017, giving it some amount of (uncharacterized) effective $r_{sb}$.

In addition to the noise characterization, the efficiency of the telescope must also be quantified. The DPFU relates flux density units incident onto the dish to equivalent degrees of thermal noise power through the following equation:

$$DPFU = \frac{\eta_A A_{geom}}{2k_B}, \tag{6}$$

where $k_B$ is the Boltzmann constant ($k_B = 1.38 \times 10^3$ Jy/K), $A_{geom}$ is the geometric area of the dish, and $\eta_A$ is the aperture efficiency. For an idealized telescope with a uniform illumination (no blockage or surface errors), the full area would be available to collect the incoming signal and the aperture efficiency would be unity. Real radio telescopes intentionally taper their illumination to minimize spillover past the primary mirror, most have secondary mirror support legs that block part of the primary aperture, and generally the surface accuracy produces a non-negligible degradation in efficiency. To determine $\eta_A$, well-focused and well-pointed observations are made of calibrator sources of known brightness, usually planets (e.g., Kutner & Ulich 1981; Mangum 1993; Baars 2007). The planet brightness temperature models from the GILDAS[115] software package were used for this calibration. For each single-dish EHT station, we determine a single DPFU value per polarization/band, except for JCMT, which has measurable temporal variations from solar heating during daytime observations. A more detailed overview of the methodology for $\eta_A$ is presented in Issaoun et al. (2017a).

We separately determine the elevation-dependent efficiency factor $\eta_{el}$ (or gain curve) due primarily to gravitational deformation of each parabolic dish. The characterization of the telescope's geometric gain curve is particularly important for the EHT, which often observes science targets at extreme elevations in order to maximize $(u, v)$ coverage. The elevation-dependent gain curve is estimated by fitting a second-order polynomial to measurements of bright calibrator sources continuously tracked over a wide range of elevation (see Figure 9 and the technical memo by Issaoun et al. 2017b). In the EHT array, SMT, PV, and APEX have characterized gain curves. The gain curve is parameterized as a second-order polynomial about the elevation at maximum efficiency:

$$\eta_{el} = 1 - B(el - el_{max})^2 . \quad (7)$$

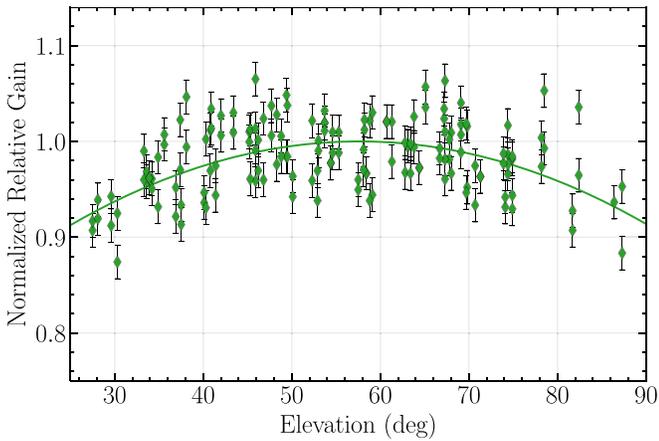

**Figure 9.** Example of a gain curve fit to single-dish normalized flux density measurements of calibrators at the SMT (Issaoun et al. 2017b).

The JCMT has no elevation dependence at 230 GHz as it is operating at the lower end of its frequency range. The LMT has an adaptive surface that is able to actively correct for surface deformation as a function of elevation. Through observations of planets, the LMT was determined to have a flat 1.3 mm gain between 25° and 80° to within 10% uncertainty. At the SPT, the elevation of extra-solar sources is constant, and therefore possible elevation-dependent efficiency losses remain uncharacterized.

We also mitigate a number of pathological issues uncovered in the 2017 data affecting the visibility amplitudes in a priori calibration. Additional loss of coherence in the signal chain at PV due to impurities in the LO, an excess noise contribution at APEX due to the inclusion of a timing signal, and the partial SMA channel dropouts were identified during data processing. Correction factors for the visibility amplitudes on baselines to these sites were estimated, as explained in the Appendix. These correction factors translate to a square multiplicative effect on the station-based SEFDs, as shown in Table 2. In the a priori calibration metadata, the multiplicative factors were folded into the DPFUs for PV and APEX and into the $T_{sys}^*$ measurements for SMA (due to its time dependence). Representative median values for the aperture efficiency, DPFU, effective system temperature, and SEFD on EHT primary targets (M87 and Sgr A*) for each station participating in the EHT 2017 observations are shown in Table 2. A site-by-site overview of the derivation of a priori calibration quantities is given in a technical memo (Janssen et al. 2019b).

### 6.1.2. Calibrating Visibility Amplitudes

The $T_{sys}^*$, DPFU, and elevation gain data for all stations are aggregated in ANTAB format text files. They are subsequently matched with observed visibilities for a given source using linear interpolation. Visibility amplitudes are calibrated in units of flux density by multiplying the normalized visibility amplitudes by the geometric mean of the derived SEFDs of

the two stations across a baseline $i$–$j$:

$$|V_{ij}| = \sqrt{SEFD_i \times SEFD_j} \, |r_{ij}|, \quad (8)$$

where $|V_{ij}|$ is then the calibrated visibility amplitude in Jy on that baseline, as in Equation (2).

Figure 10 shows the scan-averaged S/N on individual baselines, which is proportional to the phase-calibrated correlated signal, as a function of the projected baseline length (top panel), and the equivalent correlated flux density after a priori calibration (center panel) for observations of M87 (left) and 3C 279 (right) on April 11. The split in the S/N distributions is due to the difference in sensitivity between the co-located sites ALMA and APEX, leading to simultaneous baselines with two levels of sensitivity. The a priori calibration process puts all points on the same flux density scale (via Equation (8)), and the resulting data variations can thus be attributed to source structure, no longer dominated by sensitivity differences between baselines.

### 6.1.3. Single-dish Error Budget

The SEFD error budget, assuming nominal pointing and focus, is dominated by the measurement uncertainty for the DPFU (see Table 3). Depending on the source elevation, the uncertainty contribution for the elevation gain may also be non-trivial (particularly for the LMT) and adds in quadrature to the DPFU error to give the SEFD error budget. The gain curve error budget is obtained from the propagation of errors on the polynomial fit parameters in Equation (7), and is itself elevation-dependent. We assume that the uncertainty in $T_{sys}^*$ is negligible as it is the variable measured closest to the individual VLBI scans and the accuracy of the chopper method is well studied (see Section 6.1.5; Kutner 1978; Mangum 2002). The measurement uncertainties associated with pointing and focus errors are not folded into these error budget estimates as they are not easily quantifiable a priori.

For all single-dish stations, the DPFU uncertainty is estimated by the standard deviation in $\eta_A$ from a distribution of planet measurements added in quadrature to the uncertainty in the model brightness temperatures assumed for the planets. The scatter in planet measurements reflects changes in telescope performance with varying weather conditions, and thus it encompasses possible fluctuations in the mean value assumed during the observing window. An exception is the JCMT during daytime observing, where $\eta_A$ has a time dependence parametrized by a fit of a Gaussian component dip as a function of local time, described in a technical memo (Issaoun et al. 2018). The uncertainty in $\eta_A(t)$ is determined through the propagation of the errors on the fit parameters via least-squares fitting. Individual uncertainty contributions of the various components and the resulting percentage SEFD error budget for each EHT station during the 2017 April observations are listed in Table 3. Site-by-site derivations of flux density calibration uncertainties during the EHT 2017 campaign are given in Janssen et al. (2019b).

### 6.1.4. Phased-array Calibration

The phased arrays combine the total collecting area of all their dishes into one virtual telescope. This depends on precise phase alignment of the signals, with an accuracy that is captured by the

**Table 2**
Median EHT Station Sensitivities on Primary Targets during the 2017 Campaign, Assuming Nominal Pointing and Focus

| Station | Diameter in 2017 (m) | Sideband Ratio | Sideband-corrected Median $T_{sys}^*$ (K) | Aperture Efficiency $\eta_A$ | DPFU (K/Jy) | Multiplicative Mitigation Factor | Median SEFD (Jy) |
|---|---|---|---|---|---|---|---|
| APEX | 12 | ⋯ | 118 | 0.61 | 0.025 | 1.020 | 4800 |
| JCMT | 15 | 1.25 | 345 | 0.52 | 0.033[a] | ⋯ | 10500 |
| LMT | 32.5 | 1.0 | 371 | 0.28 | 0.083 | ⋯ | 4500 |
| PV | 30 | ⋯ | 226 | 0.47 | 0.12 | 3.663 | 6900 |
| SMT | 10 | ⋯ | 291 | 0.60 | 0.017 | ⋯ | 17100 |
| SPT | 6[b] | ⋯ | 118 | 0.60 | 0.0061 | ⋯ | 19300 |
| SMA6 | 14.7[c] | 1.0 | 285 | 0.75 | 0.046[d] | 1.138[e], 1.515[e] | 6400 |
| ALMA37 | 73[c] | ⋯ | 76 | 0.68 | 1.03[d] | ⋯ | 74 |

**Notes.**
[a] Nighttime value for the DPFU. The daytime DPFU includes a Gaussian component dip as function of local Hawai'i time.
[b] SPT has a 10 m dish diameter, with 6 m illuminated by receiver optics in 2017.
[c] The diameter for phased arrays reflects the sum total collecting area.
[d] DPFUs for phased arrays are determined for the full collecting areas.
[e] Applied when 6.25% and 18.75% of the SMA bandwidth was corrupted, respectively.

phasing efficiency $\eta_{ph}$ (see Appendix in Paper II)

$$\eta_{ph} = \frac{|\sum \gamma_i|^2}{(\sum |\gamma_i|)^2}. \qquad (9)$$

The phasing efficiency contributes to the aperture efficiency of the phased array, and reflects the ratio of source signal power[116] observed by the phased array, versus that observed by a perfectly phased array. The complex gains $\gamma_i$ (as in Equation (2)) are taken over all the dishes in the phased array, and have zero relative phase in the case of ideal phasing ($\eta_{ph} = 1$).

The phasing efficiency as defined above is valid when the signals being combined are optimally weighted by the effective collecting area of each antenna, $A_{i,eff} \sim 1/\text{SEFD}_i$. Then the SEFD of the phased array is

$$\text{SEFD}_{array} = \frac{1}{\eta_{ph}} \left( \sum \frac{1}{\text{SEFD}_i} \right)^{-1}. \qquad (10)$$

Both SMA and ALMA use equal weights for the formation of the sum signal. Due to their homogeneity, Equations (9) and (10) are excellent approximations.

At the SMA, the phasing efficiency $\eta_{ph}$ is estimated from self-calibrated phases to a point-source model (Young et al. 2016). Phases for each dish of the connected-element array are calculated online once per integration period, which varies in the range of 6–20 s depending on the observing conditions, and the same phases are fed back as corrective phases for beamforming the phased array. The DPFU for the individual antennas that comprise the SMA are well characterized at 0.0077 K/Jy, with $\eta_A = 0.75$, and the 6 m dishes have a flat gain curve at 230 GHz, which is near the lower end of their operating frequency range (Matsushita et al. 2006). An SEFD for each antenna is calculated from DSB $T_{sys}^*$ measurements taken regularly at the time of observing. The overall SEFD for the SMA phased array is then estimated via Equation (10).

For ALMA, both amplitude and phase gain for each dish are solved during the offline QA2 processing of interferometric ALMA data, under an assumed point-source model with known total flux. The SEFDs of individual antennas are thus determined through amplitude self-calibration, automatically accounting for system noise and efficiency factors but sensitive to errors in the source model. Because ALMA data has the additional complication of linear-to-circular conversion, the phased-sum signal SEFD is determined via the full-Stokes Jones matrix of the phased array, as computed by PolConvert (Equation (15) of Martí-Vidal et al. 2016). By convention, QA2 sensitivity tables place all phasing-related factors into the $T_{sys}^*$ component of Equation (3), allowing DPFU to assume a constant value corresponding to a single ALMA antenna. Further details are provided in Section 6.2.1 of Goddi et al. (2019).

During the EHT 2017 observations, $\eta_{ph}$ was above 0.8 for ∼80% (ALMA) and ∼90% (SMA) of the time. Poorer efficiency at both sites is associated with low elevation and increased atmospheric turbulence. At ALMA, phase corrections are calculated online by the telescope calibration system and applied to the array with a loop time of ∼18 s (Goddi et al. 2019). At the SMA, integration times at the correlator can be as short as 6 s, but longer intervals are used if needed to build S/N. The corrective phases are passed through a stabilization filter before being applied, resulting in an effective loop time of ∼12–40 s for the SMA. Phasing at both sites suffers when the atmospheric coherence timescale becomes short with respect to the loop time. To minimize the impact, both arrays are arranged in tight configurations during phased array operations.

The uncertainty on the $\eta_{ph}$ measurement at the SMA is estimated to be 5%–15%, and depends primarily on the S/N of the gain solutions. The SMA (usually with six 6 m phased) has considerably less collecting area than ALMA (usually with 37 12 m dishes phased) to use for solving phase gains. For weaker sources, the uncertainty in estimating corrective phases at the SMA and in calculating the phasing efficiency can be considerable. The assumed flux of the point-source model used to self-calibrate ALMA during QA2 has a quoted 10% systematic uncertainty in Goddi et al. (2019). The uncertainties from self-calibration and phasing are uncharacterized, therefore the uncertainty of 10% for the derived SEFD of the ALMA phased array is considered a lower limit. Errors from the use of a point-source model for M87 and 3C 279 during gain calibration are expected to be small in comparison to these values. The individual uncertainties and error budget for the phased arrays are shown in Table 3.

---

[116] It is common to see $\eta_{ph}^{1/2}$ defined as the phasing efficiency (e.g., Matthews et al. 2018), which scales with signal amplitude.

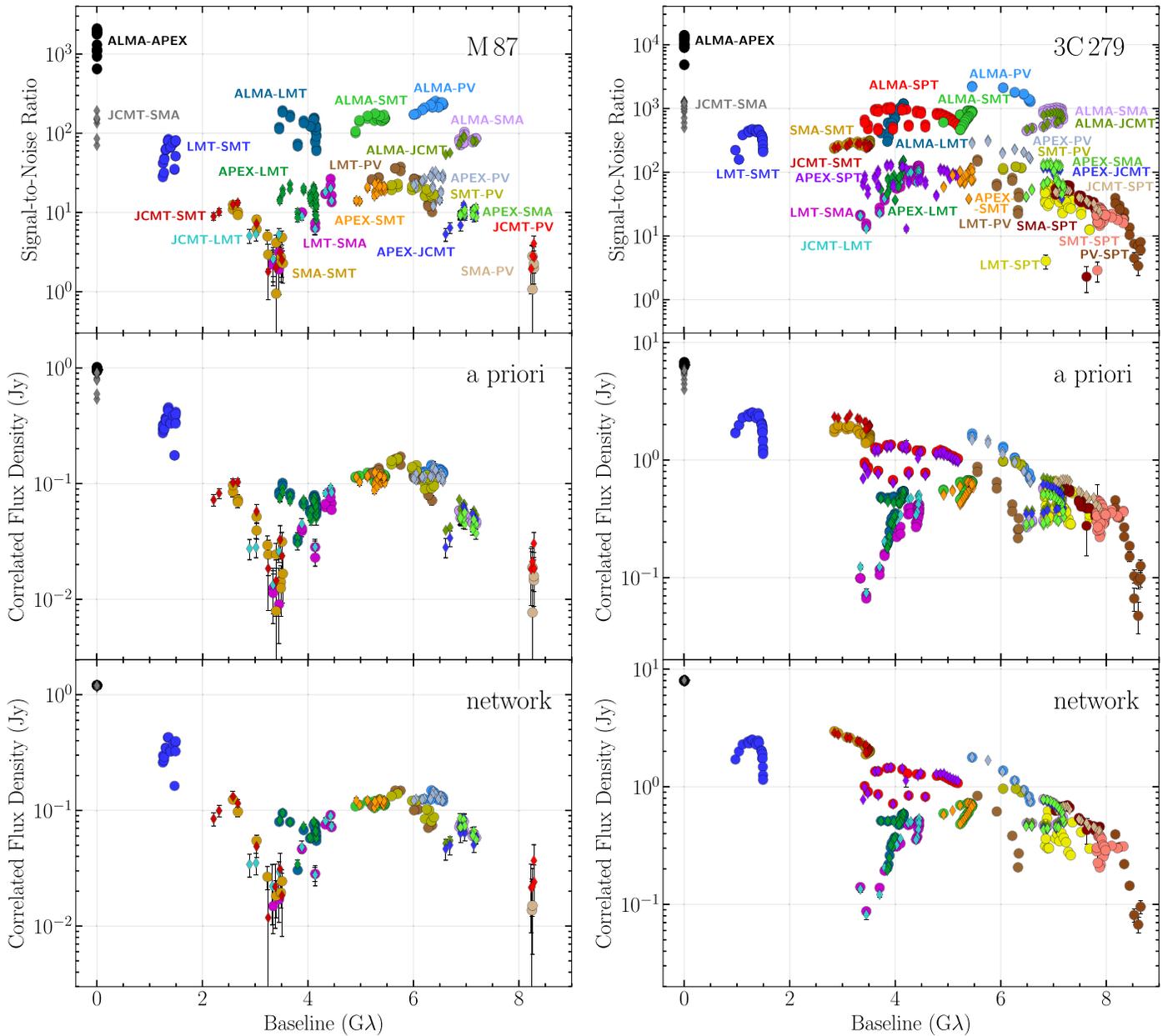

**Figure 10.** Stages of visibility amplitude calibration illustrated with the April 11 HOPS data set on M87 (left) and 3C 279 (right), as a function of projected baseline length. The two frequency bands are coherently scan-averaged separately and the final amplitudes are averaged incoherently across bands. Top: S/N of the correlated flux density component after phase calibration, both RCP and LCP. Middle: flux-density calibrated RCP and LCP values. Bottom: final, network-calibrated Stokes $I$ flux densities. Error bars denote $\pm 1\sigma$ uncertainty from thermal noise.

### 6.1.5. Limitations of a Priori Calibration

Although the DPFU is typically represented as a single value measured under good performance conditions, a station's efficiency is expected to vary with temperature, sunlight, and quality of pointing and focus. We have attempted to characterize specific time-dependent trends such as daytime dependence for the JCMT, but other factors are very difficult to decouple from the overall station behavior and associate with individual scans. Specific efficiency losses during scans, in particular due to lack of pointing/focus accuracy, are not included in the a priori amplitude calibration information for single-dish sites and remain in the underlying correlated visibilities. Therefore, the a priori error budget in Table 3 is only representative of global station performance and cannot be estimated for individual scans. In addition to a priori calibration, a list of problematic scans, where the station performance is known to be poor and the error budget is thus assumed to be undetermined, is passed on to analysis groups. These losses can be corrected in imaging and model fitting via self-calibration methods and amplitude gain modeling (Papers IV, VI).

The uncertainty in the chopper calibration is also difficult to quantify, as we do not know the true coupling of the hot load to the receiver (including spillover and reflection) and thus its effective temperature is uncertain (Kutner 1978; Jewell 2002). One of the key assumptions of the chopper method is the

**Table 3**
Station-based SEFD Percentage Error Budget during the 2017 Campaign, Assuming Stable Weather Conditions and Nominal Pointing and Focus (Subdominant Effects from $T_{\rm sys}^*$ Measurements and Sideband Ratios are not Shown)

| Station | DPFU Budget (%) | Gain Curve Budget (%) | $\eta_{\rm ph}$ Budget (%) | SEFD Budget (%) |
|---------|-----------------|------------------------|-----------------------------|------------------|
| APEX | 11 | 0.3 | ⋯ | 11 |
| JCMT | 11–14[a] | ⋯ | ⋯ | 11–14 |
| LMT | 20 | 10 | ⋯ | 22[b] |
| PV | 10 | 1.5 | ⋯ | 10 |
| SMT | 7 | 1 | ⋯ | 7 |
| SPT | 15 | ⋯ | ⋯ | 15[b] |
| SMA6 | 2 | ⋯ | 5–15[c] | 5–15 |
| ALMA37 | 10 | ⋯ | ⋯ | 10[d] |

**Notes.**

[a] The range in the budget at the JCMT is the result of a larger uncertainty in the calibration during daytime observing, due to its aperture efficiency time dependence.

[b] The error budget for SPT and LMT are lower limits due to uncharacterized losses, see Section 6.1.5.

[c] The range in the budget at the SMA is due to a larger uncertainty in the phasing for weaker sources.

[d] ALMA uncertainty is a lower limit from systematics caused by the assumed source flux density during QA2 calibration.

equivalence (to first order) of the hot load, ambient, and atmospheric temperatures, which allows for the correction of the atmospheric attenuation in the signal chain. Any deviation from this assumption in the $T_{\rm sys}^*$ measurements may introduce systematic biases. This can be partly mitigated by frequent measurements and monitoring of the DPFU under stable weather conditions and nominal telescope performance, to offset any significant scaling from temperature assumptions. The majority of stations in the EHT use a two-load (hot and cold loads) chopper method, with temperature refinement from atmospheric modeling, to measure the receiver noise temperature, and have radiometers to monitor the atmospheric opacity, which typically reduces uncertainty in the chopper calibration down to the 1% level (Jewell 2002; Mangum 2002). In contrast, the LMT and SPT used a single-load chopper method in 2017, leading to a larger error contribution estimated at the 5%–10% level minimum (Jewell 2002; Mangum 2002); with an error that grows rapidly at high line-of-sight angles.

Limitations in accuracy of the a priori calibration may also come from the cadence of DPFU and $T_{\rm sys}^*$ measurements, typically performed between scheduled VLBI scans or outside VLBI observing altogether. The changing dish performance during the VLBI observations and intra-scan atmospheric variations are not typically captured by these measurements, although frequent pointing and focus calibration is done during the observations to keep an optimal performance. Furthermore, the time cadence varies across participating stations due to different chopper calibration setups, pointing, and focus needs, and allocated time for the EHT observing campaign. It is therefore not atypical for self-calibration corrections in downstream analysis to slightly deviate from the attributed amplitude error budget. To maximize mutual coverage, many stations are pushed past their nominal operating conditions during EHT observations, such as the LMT or the JCMT in the early evening local time due to surface heating and instability, and the SPT at extremely low elevation and high winds. For those stations and conditions, we expect residual gains to deviate significantly from the a priori amplitude error budget. A more detailed discussion of a priori calibration uncertainties and limitations is given in Issaoun et al. (2017a).

### 6.2. Network Calibration

Network calibration is a framework to estimate visibility amplitude corrections at some sites by utilizing array redundancy and supplemental measurements of the total flux density of a source (Fish et al. 2011; Johnson et al. 2015; Blackburn et al. 2019). It allows for absolute amplitude calibration of intra-site baselines and tightens consistency between simultaneous baselines to co-located sites when both sites are observing (see the bottom panels of Figure 10). It makes fewer assumptions than other techniques such as self-calibration and does not assume a specific compact source model.

Network calibration makes two related assumptions. The first is that redundant baselines in the EHT array (e.g., ALMA–SMA and APEX–JCMT) share the same model visibility. The second is that co-located sites provide a zero-baseline interferometer (e.g., ALMA–APEX), with a corresponding visibility that is a positive real number equal to the total flux density $V_0$. We express the measured visibility $V_{ij}$ on a baseline between sites $i$ and $j$ as

$$V_{ij} = g_i g_j^* \mathcal{V}_{ij},\qquad(11)$$

where $\mathcal{V}_{ij}$ is the true visibility on that baseline, and $g_i$ and $g_j$ are the station-based residual gains assuming no thermal noise (the latter introduces uncertainty in the estimated gains).

Given two co-located sites $i$ and $j$, we can solve for the amplitudes of their gains using a third remote site, using the assumptions above, $\mathcal{V}_{ik} = \mathcal{V}_{jk}$ and $\mathcal{V}_{ij} = V_0$. In the absence of thermal noise,

$$|g_i| = \sqrt{\frac{V_{ij}}{V_0} \times \frac{V_{ik}}{V_{jk}}} \quad \text{and} \quad |g_j| = \sqrt{\frac{V_{ij}}{V_0} \times \frac{V_{jk}}{V_{ik}}}.\qquad(12)$$

Note that network calibration only provides gain estimates for those sites with a co-located partner.

In practice, thermal noise affects the accuracy of gains estimated using Equation (12). To optimize network calibration, we use all sets of baselines between co-located sites and distant sites and solve for the set of unknown model visibilities $\mathcal{V}_{ij}$ and station gains $g_j$ by minimizing an associated $\chi^2$. Specifically, for each solution interval, we minimize

$$\chi^2 = \sum_{i<j} \frac{|g_i g_j^* \mathcal{V}_{ij} - V_{ij}|^2}{\sigma_{ij}^2},\qquad(13)$$

where $\sigma_{ij}$ is the thermal uncertainty on $V_{ij}$. We implemented network calibration via this minimization procedure within the `eht-imaging` library (Chael et al. 2016, 2018).

For the EHT 2017 April observations, network calibration is performed on frequency-averaged visibility `UVFITS` data coherently time-averaged over 10 s solution intervals. Both parallel-hand visibility components (further referred to as RCP/LCP or RR/LL) are network-calibrated with shared gain coefficients, using the total intensity measured by the ALMA array as $V_0$ (Goddi et al. 2019). The assumed flux density values per band on each observing day are reported in Table 4 for both M87 and 3C 279. For each source, a constant flux

**Table 4**
Total Flux Density Estimates used for Network Calibration

| Source | Band | April 5 (Jy) | April 6 (Jy) | April 10 (Jy) | April 11 (Jy) |
|---|---|---|---|---|---|
| M87 | low | 1.13 | 1.14 | 1.17 | 1.21 |
| | high | 1.12 | 1.10 | 1.15 | 1.20 |
| 3C 279 | low | 8.61 | 8.57 | 7.99 | 8.01 |
| | high | 8.56 | 8.55 | 7.97 | 7.98 |

**Note.** The flux density values used for network calibration in SR1 come from the initial ALMA QA2 data release (2017 October), with a quoted uncertainty of 10%. Updated values are reported in Appendix B of Goddi et al. (2019) and are approximately 10% higher than shown here.

density is adopted per day, as both sources vary by <5% within an observation, well within the 10% flux density calibration error budget of ALMA measurements.

Network calibration enables absolute amplitude calibration of sites with a co-located partner (ALMA and APEX, SMA and JCMT) when both sites are operating, to the limit of thermal noise to the strongest remote stations. The remaining isolated sites (SMT, LMT, SPT, and PV) are unaffected by network calibration.

Following all calibration steps, Stokes $I$ total intensity components correspond to

$$V_{ij,I} = \frac{1}{2}(V_{ij,\mathrm{RR}} + V_{ij,\mathrm{LL}}). \qquad (14)$$

For JCMT, which is a single polarization station, we use the available RCP or LCP component as a proxy for the Stokes $I$ value. This corresponds to assuming zero contribution from Stokes $V$ circular polarization.

Most assumptions in the network calibration procedure are valid for all targets observed by the EHT. However, the assumption that co-located sites act as a true zero-baseline interferometer may not hold for sources with extended structure, such as M87. The distance between the SMA and the JCMT is 160 m, giving a resolution on that baseline of $\sim 1\rlap{.}''6$. The distance between ALMA (phase center) and APEX is 2.6 km, giving a resolution on that baseline of $\sim 0\rlap{.}''1$. For very compact sources, such as the quasar 3C 279, these two baselines both see point-like sources. For sources with extended structure, such as M87 and its large-scale jet, these two baselines will see slightly different structure. For example *HST*-1, a bright feature in the jet of M87 at just $0\rlap{.}''8$ from the radio core (Chang et al. 2010), produces a different response on both intra-site baselines. However, *HST*-1 has $\leqslant 1\%$ of the total core flux density of M87 as measured by ALMA (Table 4), so its effect on the network calibration gain solutions for ALMA and APEX is insignificant in comparison to the 10% uncertainty on the ALMA total flux density estimates.

## 7. Final Data Products

### 7.1. Data Release Specification

The SR1 data on M87 and 3C 279 represent a subset of a more comprehensive engineering release (ER) data production (ER5) for the EHT 2017 observations, after extensive internal validation and review. ER5 data are themselves derived from a fifth revision (Rev5) correlation data product. Information about accessing SR1 data and the software used for

analysis can be found on the Event Horizon Telescope website's data portal.[117]

The sequence of correlation and engineering releases represents a year-long effort of identifying and mitigating data issues, and developing new software and procedures; first on secondary targets for ER1–ER3 and then including EHT primary science targets for ER4–ER5. Each internal engineering data release was subject to an independent review by a panel of experts not involved in the data preparation, before being made available for downstream analysis, including imaging and model fitting. The HOPS data set was present in all engineering releases, receiving the most extensive review and internal validation. AIPS data were included in ER1 for an initial comparison to HOPS on EHT secondary targets, and in ER5 for comparisons with both HOPS and the newly added CASA data set.

The final data products at the end of the calibration and reduction pipelines provide a uniform and reliable data set for scientific analysis that has been reduced and simplified by the removal of bad data (failed observations), and after compensating for non-astrophysical systematics. The data reduction process is automated and makes only minimal assumptions about the source: (1) that the target is mostly compact, and (2) that it has known a priori large-scale structure and total flux density (e.g., from ALMA observations). The calibration of systematics is therefore limited by an inability to jointly fit source parameters along with gains, but this pathway avoids introducing any strong model assumptions during the data preparation.

In addition to the raw correlator output, three levels of successive data reduction are provided, representing the assumptions made during calibration. The first level (1) includes only the phase calibration provided during fringe fitting, after which data can be averaged. At this stage, the data represent correlation coefficients and are the most fundamental data product for the formation of closure phases and closure amplitudes. This is followed by (2) data that has been brought to a physical amplitude scale (Jy) through a priori flux density calibration, and then (3) network amplitude calibrated using a priori assumptions about large-scale source structure and total flux density. The time–frequency resolutions of the various data products are presented in Table 5, and generally exceed what is needed to capture source structure. This resolution is chosen to allow for a manageable data volume while still providing flexibility for downstream time–frequency averaging as well as the fitting of any residual systematics through additional model-dependent techniques such as self-calibration.

The SR1 data release includes products of all three fringe-fitting pipelines. The HOPS pipeline data product is designated as the primary scientific EHT data set, given the degree of vetting it has received during an iterative process of five engineering releases and a current performance advantage at low S/N. The CASA and AIPS data sets are used for validation, including direct data cross-comparisons as well as validation of downstream analysis results. Each data product is provided in UVFITS format. The choice of format was motivated by the need for common output across all pipelines, and easy loading, inspection, and imaging in all software used in the downstream analysis efforts and via readily available Python modules. A suite of metadata accompany the release,

---
[117] https://eventhorizontelescope.org/for-astronomers/data

**Table 5**
Data Products Available in SR1

| Stage | $\Delta t$ (s) | $\Delta\nu$ (MHz) | Low Band (GB) | High Band (GB) |
|---|---|---|---|---|
| Corr. Data (Rev5) | 0.4 | 0.5 | 665 | 713 |
| Phase Cal. (SR1) | 0.4 | 58.0 | 7.9 | 8.0 |
| A Priori Cal. (SR1) | 0.4 | 58.0 | 7.9 | 8.0 |
| Network Cal. (SR1) | 10.0 | 1875.0 | 0.117 | 0.121 |

**Note.** Integration time $\Delta t$ and frequency averaging windows $\Delta\nu$ are given, as well as total data volumes for low- and high-band subsets, which have slightly different coverage.

such as the ANTAB tables used for a priori calibration, documentation and validation tests for each processing and calibration stage, assessment of derived calibration solutions, and suggested flagging information from investigations of station performance.

The first science release only provides calibrated Stokes $I$ (total intensity) products for M87 and 3C 279. A summary of the data set content and S/N statistics is shown in Table 6, and a cumulative histogram of the Stokes $I$ component S/N in the fully averaged data set is shown in Figure 11. A median reported thermal uncertainty is about 7 mJy on non-ALMA baselines and, remarkably, only about 0.7 mJy on baselines to ALMA for Stokes $I$ single-band scan-averaged visibilities. In this first science release, the issue of polarimetric leakage calibration and correction is not addressed. Leakage has a relatively small influence on the total intensity and it is sufficient to parameterize the effects of leakage as a systematic source of non-closing errors (see Section 8). Future EHT results concerning polarimetry and other Stokes components will necessarily involve leakage calibration.

### 7.2. Closure Quantities

While the data release consists of reduced complex visibilities, derivative closure data products are particularly important for downstream data analysis, as well as for the description of data uncertainties. Unlike complex visibilities, closure quantities are robust against station-based gain errors. They are, however, susceptible to systematic non-closing errors, discussed in Section 8. For the needs of this Letter, we only provide brief definitions and description of conventions.

We define a closure phase formed from baseline visibilities on a closed triangle $ijk$ as

$$\psi_{C,ijk} = \mathrm{Arg}\,(V_{ij}V_{jk}V_{ki}), \tag{15}$$

with a corresponding uncertainty

$$\sigma_{\psi_{C,ijk}} \approx \sqrt{S_{ij}^{-2} + S_{jk}^{-2} + S_{ki}^{-2}}, \tag{16}$$

where $S_{ij}$ is the estimated S/N, associated with the $V_{ij}$ visibility, that is

$$S_{ij} = \frac{|V_{ij}|}{\sigma_{ij}}. \tag{17}$$

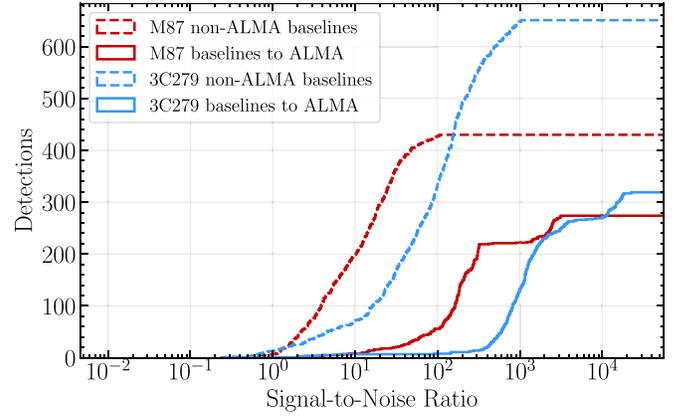

**Figure 11.** Cumulative histogram of Stokes $I$ S/N in the HOPS data set for all observations of M87 and 3C 279, using fully averaged data. Solid curves represent baselines to ALMA, while the dashed curves show all other baselines.

**Table 6**
Content of the SR1 Data Set

| | | HOPS | CASA | AIPS | Shared | Max |
|---|---|---|---|---|---|---|
| M87 | scans | 72 | 71 | 71 | 71 | 72 |
| | detections | 771 | 753 | 706 | 702 | 898 |
| | median S/N | 31.4 | 27.3 | 25.9 | ⋯ | ⋯ |
| | (shared set) | 36.6 | 31.8 | 26.4 | ⋯ | ⋯ |
| | all closure phases | 912 | 889 | 790 | 784 | 1130 |
| | (non-redundant) | 482 | 470 | 432 | 429 | 579 |
| | all closure amps | 1938 | 1890 | 1569 | 1557 | 2520 |
| | (non-redundant) | 410 | 399 | 361 | 358 | 507 |
| 3C 279 | scans | 71 | 71 | 68 | 68 | 71 |
| | detections | 954 | 937 | 972 | 913 | 1246 |
| | median S/N | 250 | 219 | 187 | ⋯ | ⋯ |
| | (shared set) | 259 | 230 | 213 | ⋯ | ⋯ |
| | all closure phases | 1313 | 1285 | 1370 | 1258 | 1918 |
| | (non-redundant) | 631 | 618 | 646 | 607 | 864 |
| | all closure amps | 3342 | 3276 | 3591 | 3207 | 5361 |
| | (non-redundant) | 560 | 547 | 578 | 536 | 793 |

**Note.** Data products in the fully averaged SR1 data set. The shared data set is composed of only those detections that are reported by all three pipelines. The max data set is a theoretical maximum calculated assuming perfect realization of the observation schedules. The full set of all closure quantities is shown, which is used to estimate systematics in Section 8; as well as the non-redundant set, which reflects the actual number of unique phase and amplitude degrees of freedom measured by the (uncalibrated) array.

Formation of closure phase cancels the station-based gain factors that appear in Equation (2). In the case of visibility amplitudes, the gain factors can be similarly canceled by the formation of the log closure amplitude, defined as

$$\ln A_{C,ijk\ell} = \ln \frac{A_{ij}A_{k\ell}}{A_{ik}A_{j\ell}}, \tag{18}$$

for a quadrangle $ijk\ell$, where "ln" is a natural logarithm and $A_{ij}$ represents debiased amplitude

$$A_{ij} = \sqrt{|V_{ij}|^2 - \sigma_{ij}^2}. \tag{19}$$

The associated uncertainty of log closure amplitude is

$$\sigma_{\ln A_C, ijk\ell} \approx \sqrt{S_{ij}^{-2} + S_{k\ell}^{-2} + S_{ik}^{-2} + S_{j\ell}^{-2}}. \quad (20)$$

Uncertainties reported in Equations (16) and (20) are calculated based on propagation of thermal visibility errors and are strictly correct in a high S/N limit, where distributions of both types of closure quantities are well approximated with a normal distribution. The number of closure quantities that can be derived from SR1 visibilities is given in Table 6. The numbers describe a fully averaged (i.e., scan and 4 GHz band-averaged) data set. We give the number of all closure quantities, corresponding to the full (or maximal) set formed from all possible loops over three or four stations in every scan. The full set has a balanced representation of baselines, and is used to estimate systematic errors in Section 8.4. Elements of a maximal set are, however, not independent (the set is highly redundant). We also provide the number of closure products in the non-redundant (or minimal) set. This is a reduced subset that captures all the available information in the closure quantities. Selection of a particular non-redundant data set is not unique and in general non-trivial (L. Blackburn et al. 2019, in preparation).

When intra-site baselines are present in the array, a special set of trivial closure quantities can be formed. Such closure phases and log closure amplitudes are zero by construction, within statistical uncertainties. While they do not carry any direct information about the source compact structure, they are useful for network calibration (Section 6.2) and the characterization of uncertainties, presented in Section 8.

### 7.3. Data Features

Certain properties of the reduced data can be directly observed in the behavior of visibilities and closure quantities. The data indicate remarkable persistent features in the structure of the M87 compact emission, as well as source structural variability on a timescale of days. In this section we give a rudimentary interpretation of these features. The implications of these basic features for the imaging, modeling, and scientific interpretation of the source structure are explored in companion Letters (Papers I, IV, V, VI).

Figure 12 shows the aggregate baseline coverage for EHT 2017 observations of M87 and 3C 279 via the HOPS pipeline. The coverage and data properties via the other two pipelines are comparable. Our shortest baselines are between co-located sites (SMA–JCMT and ALMA–APEX). These baselines are sensitive to arcsecond-scale structure, while our longest baselines are sensitive to microarcsecond-scale structure. For M87, the highest resolution (fringe spacing of 25 μas) is achieved in the east–west direction on baselines joining the Hawai'i stations to PV, while for 3C 279 the highest resolution (fringe spacing of 24 μas) is achieved in the north–south direction, on PV and SMT baselines to the SPT.

The 2017 observations led to detections on all baselines for M87. A longer averaging time (up to scan duration) is enabled by the atmospheric phase corrections performed by all three pipelines. Figure 10 (top-left panel) shows the S/N as a function of projected baseline length for M87 on April 11, for fully averaged data. A similar distribution is also shown for 3C 279 in Figure 10 (top-right panel), with around an order of magnitude difference due to the higher total flux density of 3C 279 compared to M87 (Table 4).

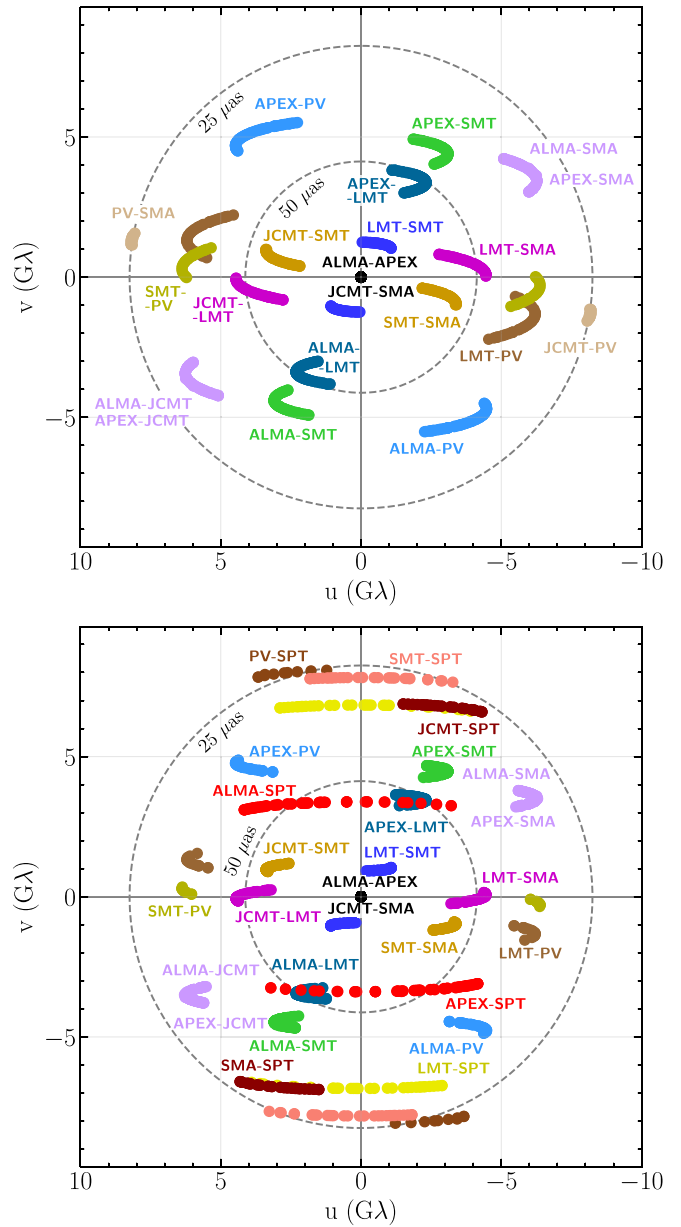

**Figure 12.** Aggregate (u, v) coverage for M87 (top panel) and 3C 279 (bottom panel) for the 2017 April observations, comparable for all three pipelines. Co-located sites (SMA/JCMT and ALMA/APEX) result in redundant baselines. The dashed circles show baseline lengths corresponding to fringe spacings of 25 and 50 μas.

The correlated flux density for M87 on April 11 after amplitude and network calibration is shown in Figure 10 (bottom left). There is a pronounced secondary peak in the visibility amplitudes with two minima on either side, interpreted as visibility nulls. The first of these nulls occurs at ∼3.4 Gλ. It is steep on the east–west oriented LMT and SMT baselines to the Hawai'i stations, and shallower on the north–south oriented ALMA and APEX baselines to LMT at the same baseline length. The second null in amplitude is observed at ∼8.3 Gλ, on the east–west oriented PV baselines to the Hawai'i stations. The correlated flux density for 3C 279 on

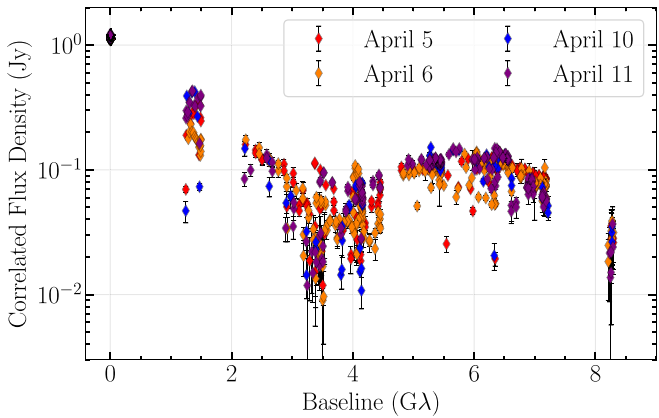

**Figure 13.** Correlated flux density of M87 as a function of projected baseline length for all four days of observations, from HOPS data that has been fully averaged. Outliers are due to reduced performance of the LMT or the JCMT. Error bars denote ±1σ uncertainty from thermal noise.

April 11 after amplitude and network calibration is also shown in Figure 10 (bottom right panel). The trend in the visibility amplitudes is clearly different from that seen in M87. 3C 279 appears to have more complex structure on long baselines, and the structure varies with baseline position angle.

### 7.3.1. Persistent Structural Features

Figure 13 shows the correlated flux density after amplitude and network calibration as a function of baseline length for all four days of observations of M87 via the HOPS pipeline. The network-calibrated amplitudes show broad consistency over different days, and are consistent between pipelines (Section 8.5). The majority of notable low-amplitude outliers across days are due to reduced efficiency of the JCMT or the LMT on a select number of scans (caused by, e.g., telescope pointing issues or surface instability). Although the amplitudes of these data points are low, closure information remains stable and is unaffected by station gain. This is shown by comparing the erratic amplitudes on the LMT–SMT baseline in Figure 13 (cluster of points at about 1 Gλ) with the smooth trends in closure phase for the ALMA–LMT–SMT triangle (Figure 14, top left) and in closure amplitude for the ALMA–LMT–APEX–SMT quadrangle (Figure 14, top right).

The secondary peak in amplitude and the location of the two nulls are persistent for all four days. These signatures in the visibility amplitudes suggest that the source is not changing dramatically over several days, is compact with a characteristic spatial scale of ≲50 μas, and exhibits similar structure over a range of baseline position angle. Long baselines with various orientations lie in a stable trend along the second peak, and a minimum in amplitude at 3.4 Gλ is seen on both the east–west and north–south oriented baselines.

While the overall trend may indicate a compact and nearly circularly symmetric structure that is stable in time, a more detailed inspection of the data set suggests the presence of a slight anisotropy, also made evident by multiple measurements of non-zero closure phase. This can be seen comparing the ALMA/APEX–LMT and SMA/JCMT–LMT amplitudes in Figure 10 (bottom left). Both baselines probe a (u, v) distance of about 3.4 Gλ, but they have a very different, nearly perpendicular orientation (Figure 12). Flux density measured on the north–south oriented ALMA–LMT baseline is a few times larger than that for the east–west oriented SMA–LMT baseline. These properties translate to striking source features in imaging and model fitting, presented in Papers IV and VI, respectively.

### 7.3.2. Time Variability

M87 was observed on the two consecutive nights of April 5/6 and again four nights later for the two consecutive nights of April 10/11. We observe clear indications of modest source evolution between the two pairs of nights, and broad consistency within each pair. The evolution can be seen particularly well in the behavior of robust closure quantities.

Across the full set of closure quantities, some closure phases formed by wide and open triangles (e.g., ALMA–LMT–SMA, Figure 14, bottom left) show different closure phase trends between the first pair of days and the second pair. Additionally, the east–west oriented LMT–SMA–SMT triangle shows different closure phase trends between the two pairs of days (Figure 14, bottom center), but the equivalent triangle in the opposite orientation, LMT–PV–SMT, shows no such trend (Figure 14, top middle).

Strong night-to-night variability of closure phases is associated with baselines probing (u, v) components close to the first visibility amplitude null, where visibility phases are particularly sensitive to small structural changes. The LMT–Hawai'i baselines are particularly affected. Rapid swings of closure phase, as large as 200° in 2 hr, are found for the LMT–SMA–SMT triangle, but exclusively for the latter pair of nights on April 10/11. Triangles that do not probe the 3.4 Gλ null location indicate less variability, e.g., ALMA–LMT–SMT or LMT–PV–SMT. Despite larger uncertainties, similar trends are seen in log closure amplitudes (right column of Figure 14). In particular, significant differences between the two pairs of nights can be seen on the ALMA–LMT–APEX–SMA quadrangle, while the ALMA–LMT–APEX–SMT quadrangle gives more consistent values.

## 8. Data Validation and Systematics

In this section, we summarize data set validation tests, performed using diagnostic tools developed in the `eat` library framework and focusing on the properties of the final network-calibrated data products. The section is structured as follows. In Section 8.1, we discuss internal consistency tests performed during the fringe-fitting stage. In Section 8.2, the accuracy of reported thermal uncertainties is tested. In Section 8.3 we investigate the robustness of data products against decoherence with increased coherent averaging time. Section 8.4 presents internal consistency tests in each pipeline and provides estimates for the magnitude of non-closing systematic errors, which become important considerations in the error budget for high S/N measurements. Finally, in Section 8.5, direct comparisons between the three pipelines are given. A more comprehensive discussion of these automated data validation procedures is given in a technical memo (Wielgus et al. 2019).

### 8.1. Fringe Validation

During fringe detection, a number of basic tests are performed on the data that check for data integrity, false fringes, and the overall self-consistency of the detected fringe solutions and measured correlation coefficients. These fringe validation tests reflect the internal validation of each pipeline, as opposed to the overall statistical validation and cross-comparisons presented in the following subsections. In addition to identifying issues with the fringe-fitting pipelines themselves, consistent review of data products throughout engineering data production played an important role in characterizing upstream issues with the data and their correlation.

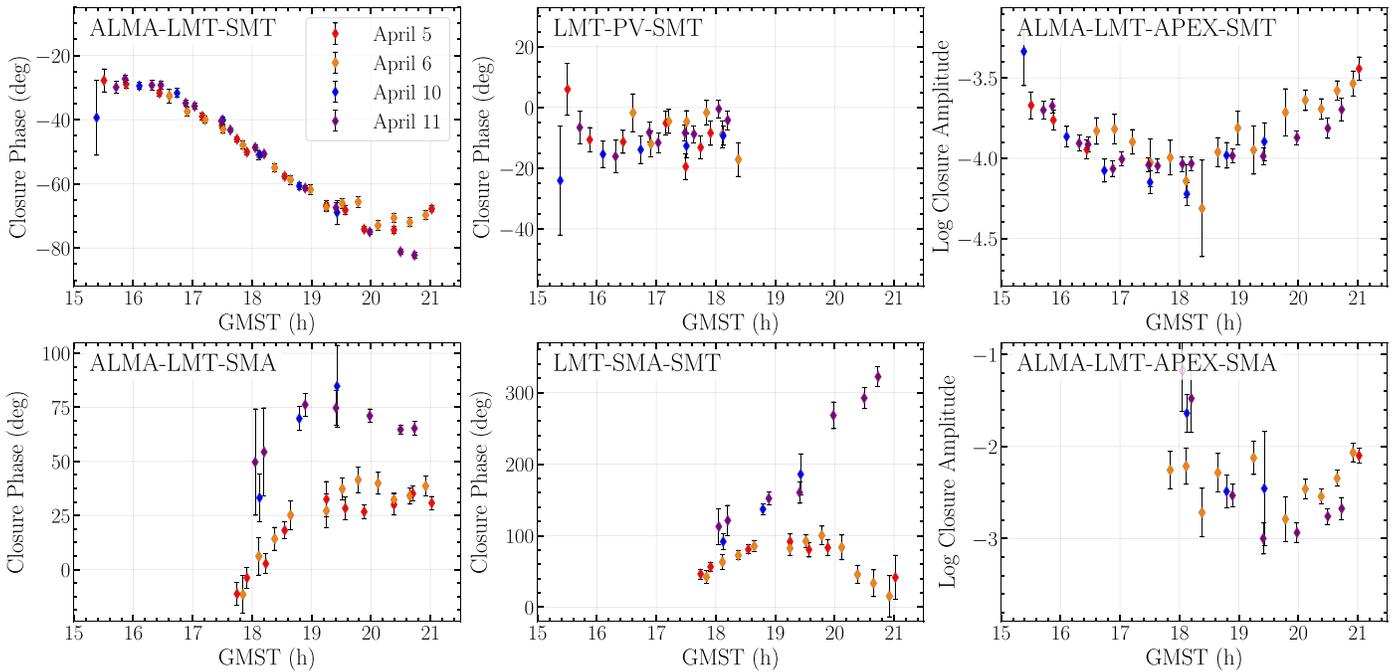

**Figure 14.** Selection of M87 closure phases (left and middle columns) and log closure amplitudes (right column) as a function of Greenwich Mean Sidereal Time (GMST) for all four observed nights from the HOPS data set. Plotted uncertainties denote $\pm 1\sigma$ ranges from thermal noise in the fully averaged data.

Figures 15 and 16 show two fringe solution consistency tests that are run as part of an automated test suite at each stage of the HOPS pipeline (Section 5.1, with details in Blackburn et al. 2019). In Figure 16, as well as in subsequent plots of distributions, the number of $3\sigma$ outliers and the size of the tested sample for each source are provided. The dashed black curve indicates a standard normal distribution with zero mean and unity variance.

The HOPS pipeline baseline-based fringe solutions (prior to the global enforcement of fringe closure) show smooth evolution across each observing night and consistency across four polarization products, which are independently fit. Delay calibration assumes a constant RCP versus LCP delay offset per night at each station, which is verified by the stability of RR−LL delays to within thermal measurement error. Independently measured delay-rates between polarizations are also consistent to within thermal error. The lack of large-deviation outliers in these fringe solution consistency tests is a strong indication that there are no false fringes or corrupted measurements above the detection threshold.

### 8.2. Thermal Error Consistency

Thermal error plays an essential role in the VLBI uncertainties, both for the visibilities as well as for the derivative closure quantities, for which uncertainties are simply propagated from the visibility errors (Section 7.2). An accurate accounting of thermal noise is essential for deriving faithful model-fitting uncertainties, and for correct noise debiasing in the case of incoherently averaged amplitudes (Rogers et al. 1995). Fundamentally, thermal uncertainty $\sigma_{th}$ in the real and imaginary components of the dimensionless complex correlation coefficient $r_{ij}$ (Equation (1)) can be estimated from first principles. Under the assumption of a stationary white noise process at each antenna

$$\sigma_{th} = \frac{1}{\eta_Q \sqrt{2 \, \Delta t \, \Delta\nu}}, \qquad (21)$$

where $\Delta t$ is the integration time, $\Delta\nu$ is the averaged bandwidth, and $\eta_Q$ is the factor that accounts for quantization efficiency. The thermal uncertainties reported by each pipeline depend on the self-consistent tracking of scale factors through data conversion and calibration, as well as accounting for the data weights and bandpass response over the averaging windows in Equation (21).

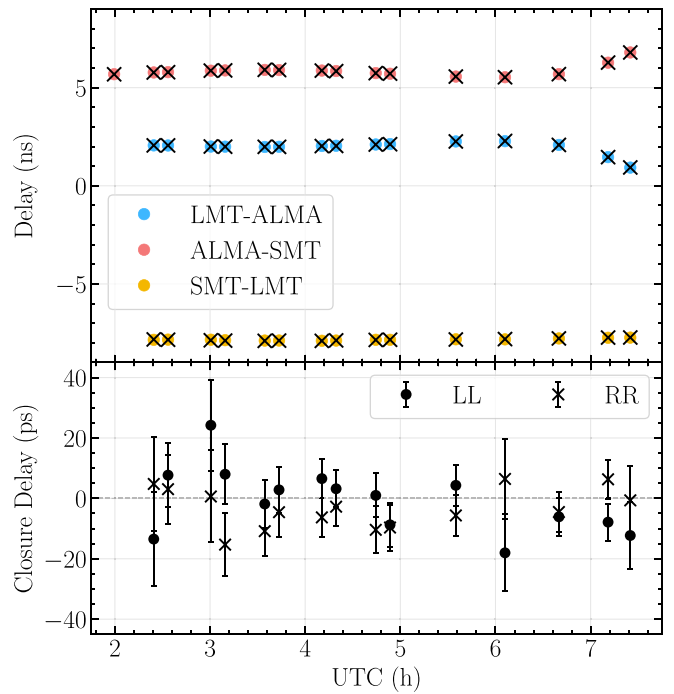

**Figure 15.** Measured residual relative delays for selected M87 baselines on April 11, reported by the HOPS pipeline (Section 5.1) prior to explicit fringe closure. The top panel shows smooth delay trends over the night for both parallel hands, LL (dots) and RR (crosses). The bottom panel shows the sum of the delays on this closed triangle, which is consistent with the expected value of zero to within statistical errors. After fringe closure, RR and LL are set to the same delay, and closure delay is zero by construction.

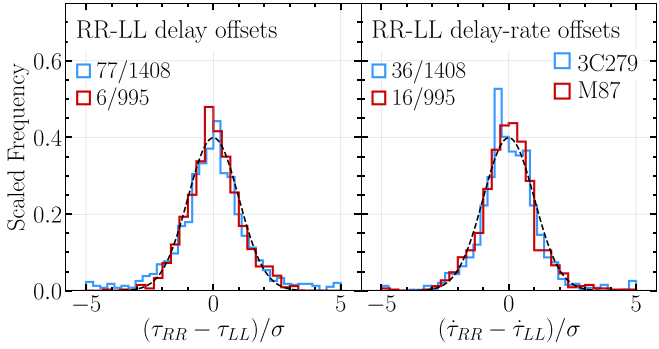

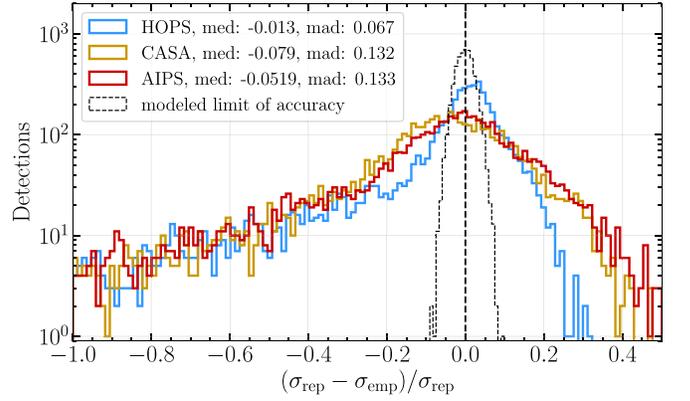

**Figure 16.** Delay and delay-rate differences between RR and LL parallel-hand fringe detections (S/N > 7) from the HOPS pipeline in units of thermal measurement uncertainty, along with the fraction of $3\sigma$ outliers. A small amount of systematic error is added in quadrature to delay (1 ps) and delay-rate (0.1 fs/s). The RR–LL differences are formed before fringe closure (after which they are zero by construction). These small differences demonstrate that there are no false fringes and that the relative difference between RCP and LCP feeds is stable at each site. Combined errors $\sigma^2 = \sigma^2_{RR} + \sigma^2_{LL}$ are used.

**Figure 17.** Joint M87 and 3C 279 histograms of differences between reported thermal uncertainties $\sigma_{rep}$, and empirically estimated uncertainties $\sigma_{emp}$. The dashed black histogram shows the limiting accuracy (high S/N, zero variance of $\sigma_{rep}$) of the empirical estimator from the finite number of 0.4 s measurements available per scan. Median (med) and median absolute deviation (mad) of each distribution are given.

The UVFITS file format formally associates a weight $w$ for each visibility measurement, with associated reported uncertainty $\sigma_{rep} \equiv 1/\sqrt{w}$. In the ideal case, $\sigma_{rep}$ properly represents thermal uncertainties, $\sigma_{rep} = \sigma_{th}$. For the HOPS and CASA pipelines, the thermal uncertainty is determined from first principles. However, the weights for the AIPS pipeline require a large scaling factor to be applied for their final output to ensure that $\sigma_{rep} = \sigma_{th}$.[118] We derive this correction factor using the scatter from differences in adjacent high-S/N visibility phases. For CASA, the direct interpretation of reported weights as $1/\sigma^2_{th}$ also leads to a small bias, resulting in underestimation of $\sigma_{th}$ by approximately 5%, as estimated by the closure phase-differencing technique.

We test the scan-by-scan accuracy of $\sigma_{rep}$ via a comparison with an empirical estimator $\sigma_{emp}$, fitting the moments of visibility amplitudes distribution. We estimate $\sigma_{emp}$ for each scan, baseline, band, and polarization combination, by using moment matching of the visibility amplitude distribution over the scan duration (Wielgus et al. 2019). Each ensemble is composed of, on average, 900 individual visibility amplitude measurements. Figure 17 shows distributions of $(\sigma_{rep} - \sigma_{emp})/\sigma_{rep}$ for all three SR1 processing pipelines, using the 5399 ensembles shared by the pipelines. The median of each distribution (med) is given in the legend of Figure 17, and shows ensemble values that are roughly consistent with the alternative closure phase differencing test. The distributions have large tails at negative values, where the empirical uncertainty exceeds the reported uncertainty. These tails are predominantly from high S/N scans with significant true intra-scan amplitude gain variation, which inflates $\sigma_{emp}$ and biases the median slightly downward. The amplitude distribution test provides a scan-by-scan estimate of the thermal error and is most reliable at low S/N; while the closure phase differencing test is appropriate at high S/N, longer integrations, and under the assumption of a constant scaling factor for $\sigma_{rep}/\sigma_{emp}$. The median absolute deviation (mad) is given as a measure of the associated uncertainties on $\sigma_{rep}$, and is fundamentally limited by the finite sample size of the estimator. From these metrics, the HOPS data set provides the most accurate accounting of thermal uncertainty.

### 8.3. Temporal Coherence after Calibration

All three data pipelines correct for changing visibility phase over scans, both in the correction for a linear drift via the delay-rate and in corrections for stochastic, station-dependent wander from atmospheric contributions (see Section 5). Although these corrections do not provide absolutely calibrated visibility phase, they eliminate differential wander on short timescales, allowing the visibilities to be coherently averaged for longer intervals than the atmospheric coherence time. An imperfect phase correction will lead to decoherence in the averages, which, in severe cases, may introduce non-closing amplitude errors.

To evaluate the performance of the phase correction algorithms, we compute two quantities for each scan: the amplitude $A_{scan}$ resulting from coherent averaging visibilities over the full scan (3–7 minutes) and subsequent debiasing (Equation (19)), and the amplitude $A_{2s}$ obtained from 2 s coherently averaged visibility segments that were subsequently incoherently averaged over the full scan (Rogers et al. 1995; Johnson et al. 2015). The ratio $A_{scan}/A_{2s}$ then quantifies the loss in amplitude from uncorrected phase fluctuations within scans.

Figure 18 shows cumulative histograms of $A_{scan}/A_{2s}$ for a common subset of 4688 ensembles (subsets of unique scan, baseline, band, and polarization) shared between pipelines, with an S/N > 7 threshold. While small errors in the estimated thermal noise have little effect on the S/N of coherent averages, they can significantly affect the outcome of incoherent averaging. Thus, only for this particular test, we applied a fixed correction factor of 1.05 to CASA thermal noise estimates $\sigma_{rep}$ before incoherent averaging, to account for the small bias in this pipeline discussed in Section 8.2. For all three pipelines, the coherence of the phase-corrected data is significantly better than that of data with no atmospheric phase correction (the gray curve in Figure 18; see also Figure 2 of Paper II). These results demonstrate that coherent averaging over scans is admissible for the SR1 data set, particularly in case of the HOPS data products.

---

[118] See AIPS Memo 103; http://www.aips.nrao.edu/aipsmemo.html.

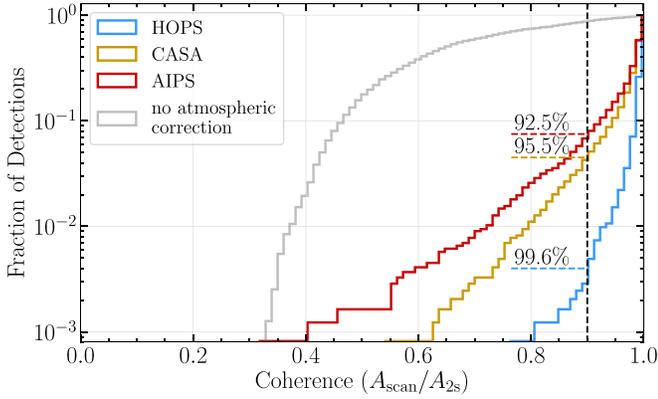

**Figure 18.** Joint M87 and 3C 279 cumulative histograms of amplitude ratios between coherent averaging for entire scans ($A_{scan}$), and coherent averaging for 2 s before incoherent averaging over scans ($A_{2s}$). The gray histogram shows the results from the HOPS pipeline with no atmospheric phase correction applied. For each pipeline, the fraction of data with coherence above 90% is indicated.

### 8.4. Intra-pipeline Validation

In this subsection we perform internal data consistency tests for each pipeline, in order to estimate the magnitude of systematic non-closing errors, e.g., related to the uncalibrated polarimetric leakage. For that purpose, we inspect closure phases and log closure amplitudes derived from the SR1 data set and evaluate consistency between (1) RR and LL components, (2) low- and high-frequency bands, and (3) trivial closure quantities. For each test, we derive a magnitude of residual errors, in excess to the reported thermal uncertainties. These values are then used to characterize the magnitude of non-closing errors in the data set, utilized in the downstream analysis.

#### 8.4.1. Quantifying Residual Errors

We evaluate the characteristic magnitude of systematic errors in the SR1 data set based on tests of distributions of closure quantities. In this approach we rely on the following modified median absolute deviation statistic:

$$\mathrm{mad}_0(Y) = 1.4826 \, \mathrm{med}(|Y|), \tag{22}$$

where "med" denotes median, the subscript zero indicates that the raw distribution moment is estimated, and the normalization factor of 1.4826 scales the result so that it acts as a robust estimator of standard deviation for a normally distributed random variable $Y$ with zero mean. We assume total uncertainties $\sigma$ associated with closure quantities to be well approximated by

$$\sigma^2 = \sigma_{th}^2 + s^2, \tag{23}$$

such that the total uncertainty consists of the known a priori thermal component $\sigma_{th}$ and a constant systematic non-closing error $s$, of unknown magnitude, added in quadrature. We then solve for the characteristic value of $s$ that enforces

$$\mathrm{mad}_0\left(\frac{X}{\sigma}\right) = \mathrm{mad}_0\left(\frac{X}{\sqrt{\sigma_{th}^2 + s^2}}\right) = 1, \tag{24}$$

where $\sigma$ is the total uncertainty associated with $X$. As an example, for RR–LL consistency of closure phases we have

$$X = \psi_{C,RR} - \psi_{C,LL},$$
$$\sigma^2 = \sigma_{\psi_{C,RR}}^2 + \sigma_{\psi_{C,LL}}^2 + s^2. \tag{25}$$

We exclude low S/N data (S/N < 7), for which the normal distribution approximation does not hold well.

#### 8.4.2. RR–LL Consistency

Consistency of closure quantities derived from RR and LL visibilities, matched for the same scan, baseline, and band, are expected to be dominated by effects related to polarimetric leakage, which remains uncalibrated in SR1 data. Assuming that some amount of leaked polarized signal mixes randomly into the parallel-hand visibilities, the degree of systematic error can be crudely approximated as

$$\delta_{leak} \approx \sqrt{2n} \, |D| |\breve{m}| < 0.14 \sqrt{n} \, |\breve{m}|, \tag{26}$$

where the number of baselines $n$ is 3 for closure phases and 4 for closure amplitudes, $|D| < 0.1$ is a leakage D-term magnitude, and $|\breve{m}|$ is a typical fractional interferometric baseline polarization (i.e., fractional linearly polarized correlated flux density relative to total intensity); see Johnson et al. (2015). If a characteristic $|\breve{m}| < 0.2$ is assumed, these upper bounds translate under Equation (26) to <2°.8 for the closure phase systematic uncertainty and <5.7% for the closure amplitude uncertainty. The results of the SR1 errors estimation by normalizing $\mathrm{mad}_0$ are summarized in Table 7. The estimated errors are consistent with the simple upper limit given by Equation (26) and roughly consistent between all data reduction pipelines. While for the high S/N source 3C 279 the leakage related errors may dominate over the thermal errors, they remain strongly subthermal for M87.

#### 8.4.3. Frequency Bands Consistency

Comparisons between low-/high-frequency bands may reveal the presence of band-specific systematics, including frequency-dependent polarimetric leakage. Apart from those, source spatial structure and spectral index both may add a small contribution. The estimated magnitudes of systematic errors found for closure phases and log closure amplitudes are given in Table 7. For all pipelines, the magnitude of characteristic closure phase inconsistency was found to be about 0.5 times the thermal uncertainty for M87 and about 1.5 times the thermal uncertainty for 3C 279 (scan-average, single-band/polarization). For 3C 279 systematic uncertainties strongly dominate over the thermal scatter, and this should be taken into account before the direct averaging of frequency bands.

**Table 7**
Systematic Errors in SR1 Data Set

| Test | M87 | | | 3C 279 | | |
|---|---|---|---|---|---|---|
| | HOPS | CASA | AIPS | HOPS | CASA | AIPS |
| RR−LL closure phases (deg) | <1.0(0.2) | <1.0(0.2) | <1.0(0.2) | 1.9(1.1) | 1.9(1.1) | 2.1(1.2) |
| RR−LL log closure amplitudes (%) | <2.0(0.2) | <3.0(0.3) | <2.0(0.2) | 3.1(1.0) | 3.6(1.2) | 3.3(1.0) |
| Stokes $I$ closure phase low/high (deg) | 1.4(0.4) | 2.5(0.6) | 2.6(0.6) | 2.2(1.5) | 2.3(1.5) | 2.0(1.3) |
| Stokes $I$ log closure amplitude low/high (%) | 5.6(0.8) | ⋯ | <10.0(1.3) | 4.5(1.8) | 5.4(2.3) | 4.8(1.8) |
| Stokes $I$ trivial closure phases (deg) | 3.7(1.1) | 2.6(0.8) | 3.2(1.0) | 1.2(1.9) | 1.0(1.5) | 1.0(1.4) |
| Stokes $I$ trivial log closure amplitudes (%) | 2.6(0.4) | 5.6(0.7) | 7.7(0.9) | 3.8(2.0) | 3.8(1.9) | 3.3(1.6) |

**Note.** Characteristic magnitudes of systematic errors, estimated using the subset of data shared by all three pipelines. Scan-averaged single-band data. Numbers in parentheses represent characteristic systematic errors in units of thermal noise.

#### 8.4.4. Trivial Closure Quantities

The intra-site baselines ALMA–APEX and JCMT–SMA provide the EHT array with multiple "trivial" closure triangles and quadrangles. Ideally, these trivial closure phases and trivial log closure amplitudes should be equal to zero, but this is not precisely true in the presence of polarimetric leakage. Furthermore, the small but finite length of intra-site baselines leads to measurements that are susceptible to contamination from large-scale structure, breaking the assumptions of a trivial closure quantity. This particular aspect is a concern for M87 and its large-scale jet. The estimated characteristic magnitude of systematic errors in trivial closure phases is given in Table 7. While for 3C 279 the magnitude of about 1° can be fully explained by polarimetric leakage, M87 systematics are inconsistent with limits given by Equation (26), suggesting the presence of an additional source of error. We illustrate the systematic-error fitting procedure in Figure 19, in which 3C 279 trivial closure phase distribution is shown, before and after adding the systematics, and is estimated to be about 1° consistently for all processing pipelines.

#### 8.4.5. Systematic Error Budget

Based on values reported in Table 7, we conclude that, for a single band, systematic errors of 3C 279 measurements can be approximated by characteristic values of about 1°.5 for closure phases and 0.03 for log closure amplitudes. For M87, leakage is not nearly as important, and other subtle effects like polarimetric calibration uncertainties may influence the total systematic error budget. Suggested systematics are 2° for closure phases and 0.04 for log closure amplitudes. For each test of closure phases and log closure amplitudes summarized in Table 7, we show related distributions in Figure 20. Errors in Figure 20 were inflated according to the above recommendation for systematic errors. A standard (zero mean, unit variance) normal distribution is shown with a dashed line. The match between the empirical distributions and the normal distribution indicates that the addition of the systematic uncertainties allows for the approximate capture of the total data uncertainty. Under the assumption of independent baseline errors, the closure uncertainties given in this section can be translated to 2% non-closing systematic uncertainties in visibility amplitudes and 1° of non-closing systematic uncertainties in visibility phases.

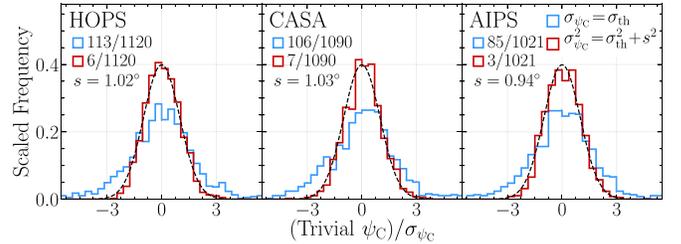

**Figure 19.** Normalized distributions of trivial closure phases for 3C 279 in three data reduction pipelines, before (blue) and after (red) accounting for the residual systematic uncertainties. Numbers indicate the fraction of $3\sigma$ outliers.

### 8.5. Inter-pipeline Consistency

Direct comparisons between corresponding data products delivered by separate pipelines allow us to quantify the degree of confidence that we may have in their properties and their dependence on specific choices in calibration procedure. Figure 21 (top) shows the distribution of visibility amplitude differences between the reduction pipelines, in units of their thermal uncertainty. Thermal errors represent a particular scale of interest; however, visibilities reduced by separate pipelines are not independent variables and share the same thermal noise realization. Another useful quantity is the relative absolute amplitude difference. As indicated in Table 8, the median relative difference between the most consistent pair of pipelines, HOPS–CASA, is 3.8%, well within the budget of a priori flux density calibration (Section 6). While for 3C 279 all three pairs represent a similar level of consistency, for M87 the HOPS–CASA pair is by far the most consistent one, as indicated in Table 8. This result is consistent with known difficulties in the processing of low S/N data with the AIPS pipeline, originating from the lack of S/N to constrain a fringe solution in the two-second intervals used for fringe fitting (Section 5.3). Distributions of differences between amplitude data products are unbiased; however, significant tails are present, with 10% of the M87 visibility amplitude data inconsistent by more than 22.8% for the most consistent pair, HOPS–CASA.

In Figure 22 we show HOPS–CASA and HOPS–AIPS scatter plots of correlation coefficient amplitude $|r_{ij}|$. The three pipelines demonstrate increasing levels of consistency at high S/N. AIPS shows a tendency to occasionally overestimate amplitude at low S/N, sometimes by a large factor, indicating a degree of over-tuning and acceptance of possible false fringes.

Contrary to visibility amplitudes, the distributions of closure phase and closure amplitude differences, shown in Figure 21, generally exhibit a spread at or below the level of thermal

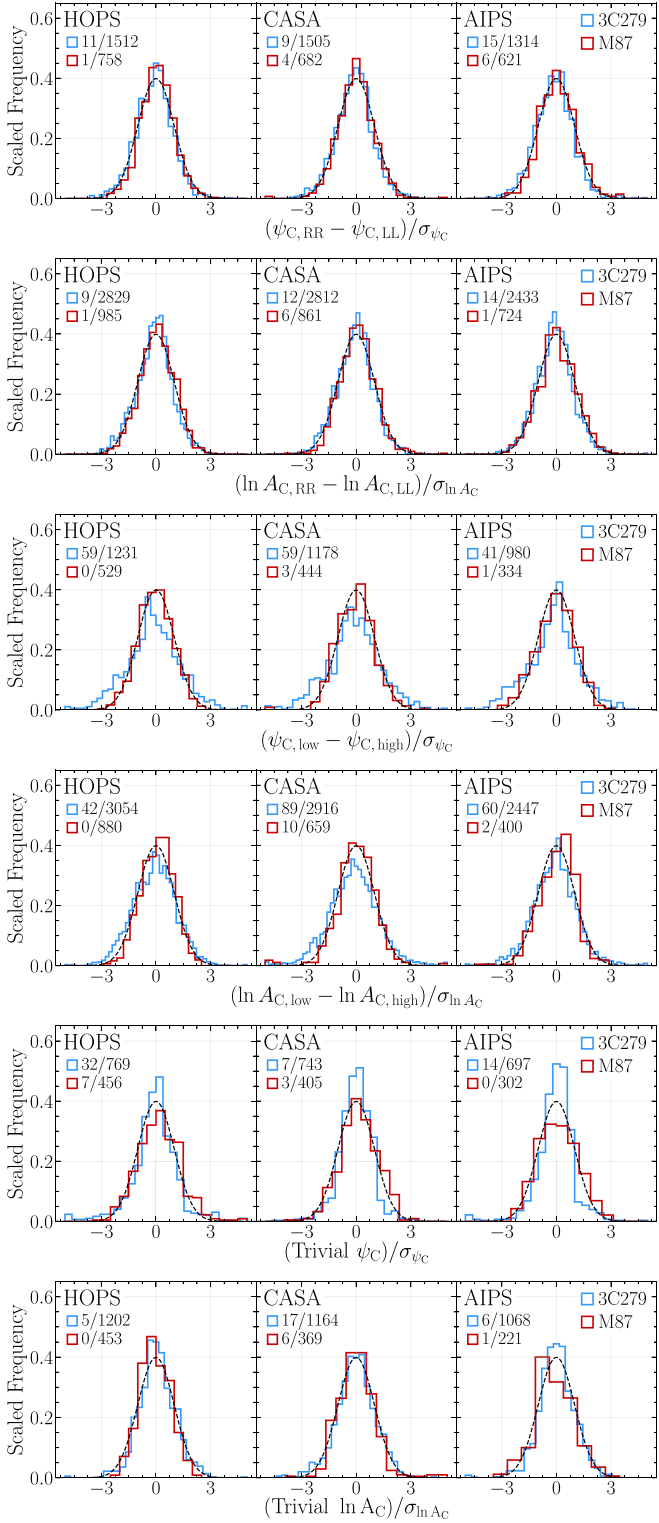

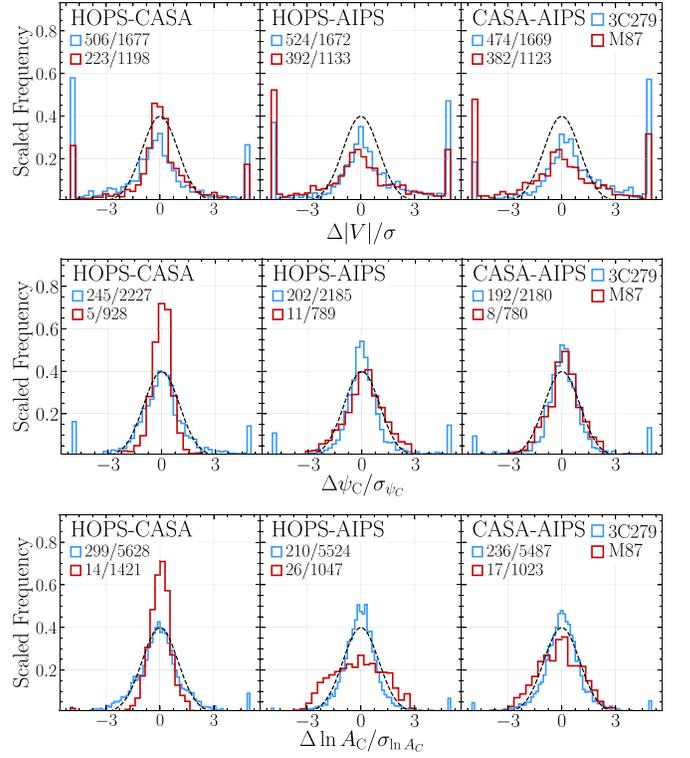

**Figure 21.** Consistency of visibility amplitudes (top), closure phases (middle), and log closure amplitudes (bottom) between the three reduction pipelines. Scan-averaged single-band Stokes $I$ data are used.

highlights the robustness of the closure quantities, independent of station-based gains.

Examples of closure phases for all three pipelines, for some of the triangles discussed in Section 7, are shown in Figure 23. While there is a broad consistency, HOPS is unique in reconstructing well-behaved closure phases on triangles including the LMT–SMA baseline over the full range of observations on April 11. To corroborate smooth trends and large closure phase evolution for these data, in two panels in Figure 23 we show data from a redundant JCMT triangle (JCMT and SMA are collocated). The redundant JCMT triangles show closure phases consistent with their SMA counterparts, and are more consistently reconstructed across the pipelines.

A bias toward zero closure phase can be seen when data are averaged in time, particularly for the AIPS data set. This is due to use of a point-source model during global fringe fitting on short time intervals (2 s for AIPS). While the individual fringe solution phases are station-based and separately close, the process biases baseline phases to zero, and closure phases generated from baseline phases averaged over multiple segments will be biased toward the point-source model. This bias is not expected in HOPS products, as HOPS fringe solutions are baseline-based and assume no structure phase for the coherent stacking of data from multiple baselines. The median bias toward zero closure phase, estimated from high S/N data at least $3\sigma$ away from zero, is about $1°$ for AIPS and CASA with respect to unbiased HOPS. However, while 90% of CASA data are biased by less than $4°.9$, 10% of AIPS data are biased by more than $8°.7$. See Wielgus et al. (2019) for an additional discussion of pipeline comparisons and associated systematics.

**Figure 20.** Closure statistics distributions after inflating errors by the amount of non-closing systematics recommended in Section 8.4.5. The plots follow the same order as the tests reported in Table 7. The dashed lines represent a standard normal distribution, and numbers show the fraction of $3\sigma$ outliers. Combined errors are used where appropriate.

uncertainty, particularly for the HOPS–CASA pair. No significant tails are present and 90% of the M87 data remain consistent to within 0.9 standard deviations of the combined thermal error budget for HOPS–CASA (Table 8). This

**Table 8**
Inter-pipeline Consistency of the SR1 Data Set

| | M87 | | | 3C 279 | | |
|---|---|---|---|---|---|---|
| | HOPS-CASA | HOPS-AIPS | CASA-AIPS | HOPS-CASA | HOPS-AIPS | CASA-AIPS |
| Median visibility error (%) | 3.8(0.7) | 7.9(1.5) | 9.4(1.5) | 1.1(1.2) | 1.2(1.3) | 1.2(1.2) |
| 90th percentile visibility error (%) | 22.8(6.0) | 52.9(7.4) | 58.3(9.5) | 5.7(9.2) | 6.7(10.0) | 7.2(8.8) |
| Median closure phase error (deg) | 3.1(0.3) | 6.8(0.7) | 6.2(0.6) | 1.4(0.7) | 1.0(0.5) | 1.0(0.6) |
| 90th percentile closure phase error (deg) | 17.7(0.9) | 39.4(1.9) | 36.5(1.7) | 6.4(3.1) | 6.0(2.5) | 5.8(2.5) |
| Median log closure amplitude error | 0.1(0.3) | 0.3(0.9) | 0.3(0.7) | 0.04(0.7) | 0.03(0.6) | 0.03(0.6) |
| 90th percentile log closure amplitude error | 0.5(1.0) | 1.4(2.4) | 1.2(2.2) | 0.15(2.3) | 0.13(1.7) | 0.13(1.9) |

**Note.** Results given for scan-averaged single-band Stokes $I$ data. Numbers in parentheses are given in thermal error units. The subset of data shared by all pipelines was used.

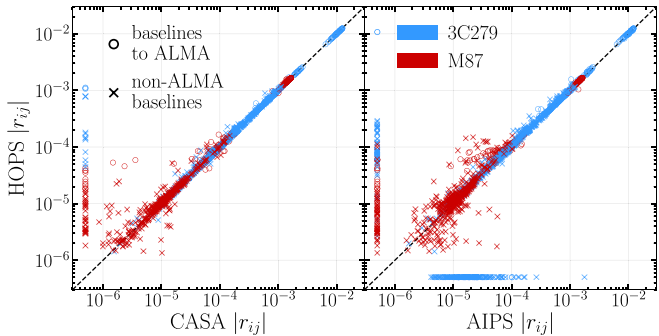

**Figure 22.** Scatter plots of complex correlation coefficient amplitudes for HOPS–CASA and HOPS–AIPS pairs of pipelines. Data are fully averaged, with an S/N > 1 threshold applied. For each detection, the mean $r_{ij}$ of available RCP and LCP components in the low and high band is given. Detections only present in one of the pipelines are shown with a fixed value of $5 \times 10^{-7}$ for the missing pipeline, and in some cases represent differences in the construction of a priori flags and fringe rejection strategies.

The HOPS pipeline benefited from a long period of development, extensive review, and internal validation through the suite of five engineering releases spanning a year-long data processing and calibration effort. In contrast, the AIPS pipeline has been used in two data releases as a secondary data set and the CASA pipeline, which is under active development, has recently been brought to maturity and included in ER5. Nonetheless, inter-pipeline comparisons of HOPS, CASA, and AIPS show a high degree of general consistency. The HOPS pipeline product was chosen as the primary science data set for SR1, based on the long validation history, level of calibration quality presented in this section, and to select a single data set for the preparation of scientific results. The other two pipelines are included in SR1 as supporting data sets for calibration, direct data comparisons, and as an independent pathway for validating the products of downstream analysis.

## 9. Conclusions

Observations from the EHT's 2017 April campaign are the first ever to have the necessary sensitivity, coverage, and resolution for horizon-scale imaging of black hole candidates M87 and Sgr A*. We have presented the complete data processing pathway that led to the first science release data set from the campaign, which includes the primary science target M87 and the secondary target 3C 279. The 2017 observations reflected a dramatic expansion of the EHT from previous years to a total of eight sites, and include for the first time ALMA as a phased array. While much more powerful, the expanded network represented a unique analysis challenge in terms of the heterogeneous nature of the array: basic telescope characteristics, weather, sensitivity, site-specific data issues, sampling rate, and channelization; and a challenge in terms of raw data volume and the needs for a homogeneous and systematized calibration strategy.

The development of processing pipelines and characterization of the data occurred over a series of five internal engineering releases, during which site-specific data issues were identified and mitigated in correlation and post-processing. SR1 is the first science release of calibrated data products arising from the mature reduction pipelines, following a series of independent internal reviews. The science data were produced without making assumptions about the detailed compact structure of the targets, and thus provide an unbiased data set for downstream imaging and modeling.

We have developed three independent processing pipelines for the initial fringe detection, phase calibration, and reduction of EHT data. The pipelines built HOPS, which has been continually developed and used for early EHT analysis over the previous decade; AIPS, the standard calibration environment for VLBI data from major facilities such as the VLBA; and CASA, a modern environment for radio interferometer calibration and analysis that has recently been augmented with VLBI capabilities. The output from each pipeline was subjected to a suite of validation tests covering self-consistency over bands and polarizations, and consistency of trivial closure quantities.

From these tests, we estimated the residual non-closing systematic errors after calibration. For M87 such errors were smaller than Stokes $I$ data thermal uncertainties even after full scan and frequency band averaging. Non-closing errors are no larger than 2° for closure phases and 4% for closure amplitudes. For 3C 279, systematics are small in an absolute sense, but they dominate the total uncertainties of the averaged data set due to the high S/N. Differences between pipelines, particularly for the robust closure quantities, were found to be largely within the total budget of uncertainties. The HOPS data were selected as the primary data set for the scientific conclusions presented in companion Letters (Papers I, IV, V, VI) with the remaining two data sets available for direct data comparisons and the cross-validation of downstream analysis.

At EHT frequencies, absolute flux density calibration is particularly challenging due to the large and time-varying 1.3 mm opacity from atmospheric water vapor, and difficulties maintaining pointing and surface accuracy particularly at the larger dishes. We have outlined the gathering and unified interpretation of auxiliary calibration data from the various sites for the purposes of a priori flux density calibration, and a strategy for estimating the residual flux density error budget within the limitations of single-dish calibration. Where available, we have made use of network redundancy to further constrain flux density calibration given generic model-independent assumptions about the source.

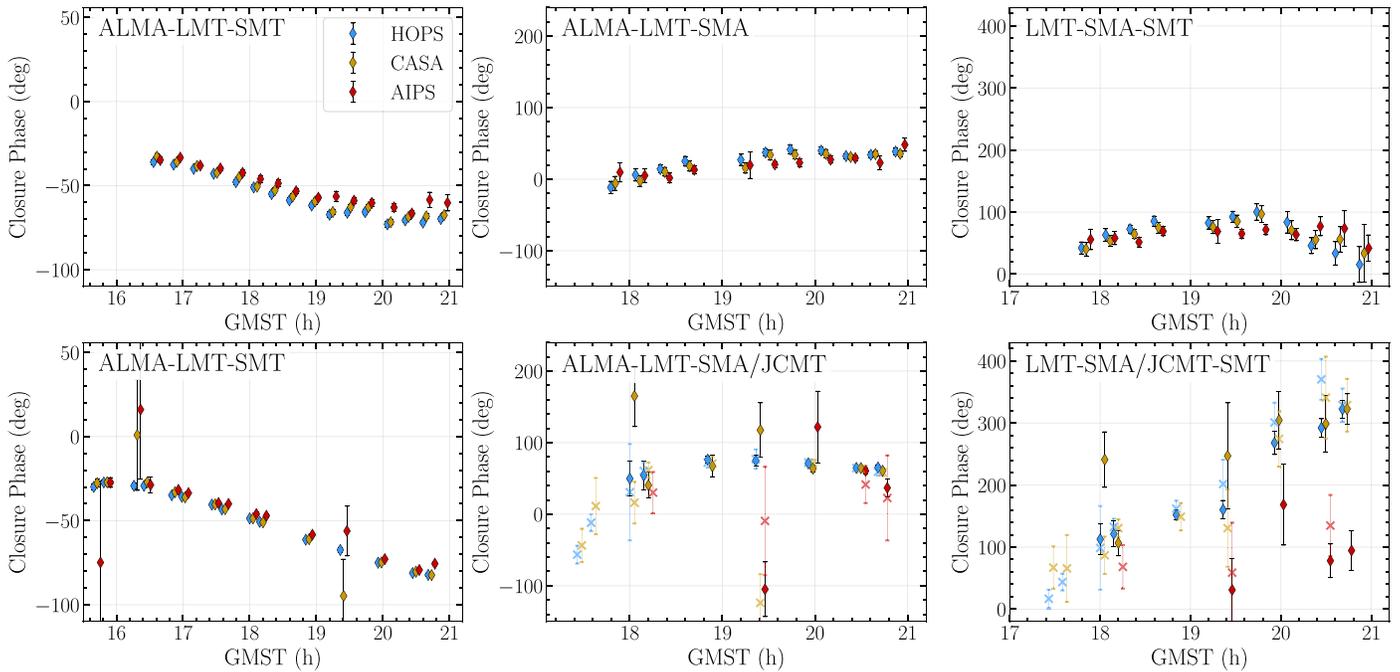

**Figure 23.** Comparison of M87 closure phases between the three fringe-fitting pipelines for selected triangles. April 6 is shown in the top row, April 11 in the bottom row. The pipelines are offset slightly in time for clarity (HOPS −3 minutes, CASA at the original timestamp, AIPS +3 minutes). Plotted uncertainties denote ±1σ ranges from thermal noise in the fully averaged data set. For the two Hawaiʻi triangles that demonstrate pronounced evolution on April 11 (see also Figure 14, bottom panels), we also include the corresponding redundant triangles with JCMT (which joined the array two scans earlier) as light crosses.

A number of salient features became apparent in the M87 data set after processing and calibration. The visibility amplitudes as a function of projected baseline length persistently show a prominent secondary peak bracketed by two nulls, the first at ∼3.4 Gλ and the second at ∼8.3 Gλ, across all four observed days. The visibility amplitudes exhibit characteristics of a compact source with a spatial scale ≲50 μas, and broad circular symmetry broken on baselines probing the first null. This spatial scale corresponds to only a few Schwarzschild radii for a ∼6.5 × 10⁹ $M_\odot$ black hole (Paper VI) at the distance of M87 (Blakeslee et al. 2009; Gebhardt et al. 2011; Cantiello et al. 2018). M87 closure phases on select triangles show clear time evolution between the two pairs of days, April 5/6 and April 10/11, providing evidence for intrinsic evolution of the source. The triangles with the largest closure phase variations between the two pairs of days have a baseline probing the (u, v) plane region about the first minimum in visibility amplitude. Analysis and interpretation of these features are presented in companion Letters (Paper I, IV, V, VI).

Although previous observations of M87 from early EHT campaigns (in 2009 and 2012) probed scales of a few tens of microarcseconds, the visibility amplitude behavior on the few baselines present remained consistent with a Gaussian source, showing no apparent finer structure (Doeleman et al. 2012; Akiyama et al. 2015). The first M87 closure phases at 1.3 mm reported in Akiyama et al. (2015) were consistent with zero to within 2σ. In addition to a first reported measurement of 1.3 mm closure amplitudes, the 2017 observations of M87 are the first to show non-Gaussian structure in the compact source and significantly non-zero closure phases.

The SR1 data provide the first opportunity for total intensity imaging of M87 (Paper IV). Efforts to characterize and remove polarization leakage are ongoing and will enable studies of the linear polarization structure of M87 and other EHT targets. Additional work to better calibrate in the presence of intrinsic source variability, as well as increased amplitude gain variability, is necessary for Sgr A* and other low-elevation targets.

For 2018, the EHT was joined by the Greenland Telescope, greatly expanding the coverage for northern sources such as M87. In the near future, the array will also be joined by the Kitt Peak 12 m telescope in Arizona and the Northern Extended Millimeter Array (NOEMA) at the Plateau de Bure observatory in France. In addition to generally improved baseline coverage, both sites provide short baselines and associated redundancy (with SMT and PV, respectively) for the array—which is particularly beneficial for amplitude calibration. The EHT doubled recorded bandwidth to a rate of 64 Gbps in 2018 as well, over four 2 GHz bands. Additional development to enable coherent fringe fitting and atmospheric phase correction across all four bands will allow the EHT to better resolve features on long baselines, short timescales, and near visibility nulls, and it will increase robustness of the array against poor weather and the potential loss of sensitive central anchor stations.

While continuous development of the instrument and the data reduction pipeline will yield future observations with improved (u, v) coverage, higher S/N, and sharper resolution, the observations carried out in 2017 already deliver data of unprecedented scientific quality. The dramatic difference between the 2017 observations and early EHT campaigns in number of participating stations, S/N, coverage, and weather conditions make the EHT 2017 data set an exceptional opportunity for scientific discoveries via, e.g., imaging and model fitting well beyond previous EHT capabilities.

The authors of this Letter thank the following organizations and programs: the Academy of Finland (projects 274477, 284495, 312496); the Advanced European Network of E-infrastructures for Astronomy with the SKA (AENEAS) project, supported by the European Commission Framework Programme Horizon 2020 Research and Innovation action under grant agreement 731016; the Alexander von Humboldt Stiftung; the Black Hole Initiative at Harvard University, through a grant (60477) from the John Templeton Foundation; the China Scholarship Council; Comisión Nacional de Investigación Científica y Tecnológica (CONICYT, Chile, via PIA ACT172033, Fondecyt 1171506, BASAL AFB-170002, ALMA-conicyt 31140007); Consejo Nacional de Ciencia y Tecnología (CONACYT, Mexico, projects 104497, 275201, 279006, 281692); the Delaney Family via the Delaney Family John A. Wheeler Chair at Perimeter Institute; Dirección General de Asuntos del Personal Académico—Universidad Nacional Autónoma de México (DGAPA—UNAM, project IN112417); the European Research Council Synergy Grant "BlackHoleCam: Imaging the Event Horizon of Black Holes" (grant 610058); the Generalitat Valenciana postdoctoral grant APOSTD/2018/177; the Gordon and Betty Moore Foundation (grants GBMF-3561, GBMF-5278); the Istituto Nazionale di Fisica Nucleare (INFN) sezione di Napoli, iniziative specifiche TEONGRAV; the International Max Planck Research School for Astronomy and Astrophysics at the Universities of Bonn and Cologne; the Jansky Fellowship program of the National Radio Astronomy Observatory (NRAO); the Japanese Government (Monbukagakusho: MEXT) Scholarship; the Japan Society for the Promotion of Science (JSPS) Grant-in-Aid for JSPS Research Fellowship (JP17J08829); JSPS Overseas Research Fellowships; the Key Research Program of Frontier Sciences, Chinese Academy of Sciences (CAS, grants QYZDJ-SSW-SLH057, QYZDJ-SSW-SYS008); the Leverhulme Trust Early Career Research Fellowship; the Max-Planck-Gesellschaft (MPG); the Max Planck Partner Group of the MPG and the CAS; the MEXT/JSPS KAKENHI (grants 18KK0090, JP18K13594, JP18K03656, JP18H03721, 18K03709, 18H01245, 25120007); the MIT International Science and Technology Initiatives (MISTI) Funds; the Ministry of Science and Technology (MOST) of Taiwan (105-2112-M-001-025-MY3, 106-2112-M-001-011, 106-2119-M-001-027, 107-2119-M-001-017, 107-2119-M-001-020, and 107-2119-M-110-005); the National Aeronautics and Space Administration (NASA, Fermi Guest Investigator grant 80NSSC17K0649); the National Institute of Natural Sciences (NINS) of Japan; the National Key Research and Development Program of China (grant 2016YFA0400704, 2016YFA0400702); the National Science Foundation (NSF, grants AST-0096454, AST-0352953, AST-0521233, AST-0705062, AST-0905844, AST-0922984, AST-1126433, AST-1140030, DGE-1144085, AST-1207704, AST-1207730, AST-1207752, MRI-1228509, OPP-1248097, AST-1310896, AST-1312651, AST-1337663, AST-1440254, AST-1555365, AST-1715061, AST-1614868, AST-1615796, AST-1716327, OISE-1743747, AST-1816420); the Natural Science Foundation of China (grants 11573051, 11633006, 11650110427, 10625314, 11721303, 11725312, 11873028, 11873073, U1531245, 11473010); the Natural Sciences and Engineering Research Council of Canada (NSERC, including a Discovery Grant and the NSERC Alexander Graham Bell Canada Graduate Scholarships-Doctoral Program); the National Youth Thousand Talents Program of China; the National Research Foundation of Korea (grant 2015-R1D1A1A01056807, the Global PhD Fellowship Grant: NRF-2015H1A2A1033752, and the Korea Research Fellowship Program: NRF-2015H1D3A1066561); the Netherlands Organization for Scientific Research (NWO) VICI award (grant 639.043.513) and Spinoza Prize SPI 78-409; the New Scientific Frontiers with Precision Radio Interferometry Fellowship awarded by the South African Radio Astronomy Observatory (SARAO), which is a facility of the National Research Foundation (NRF), an agency of the Department of Science and Technology (DST) of South Africa; the Onsala Space Observatory (OSO) national infrastructure, for the provisioning of its facilities/observational support (OSO receives funding through the Swedish Research Council under grant 2017-00648) the Perimeter Institute for Theoretical Physics (research at Perimeter Institute is supported by the Government of Canada through the Department of Innovation, Science and Economic Development Canada and by the Province of Ontario through the Ministry of Economic Development, Job Creation and Trade); the Russian Science Foundation (grant 17-12-01029); the Spanish Ministerio de Economía y Competitividad (grants AYA2015-63939-C2-1-P, AYA2016-80889-P); the State Agency for Research of the Spanish MCIU through the "Center of Excellence Severo Ochoa" award for the Instituto de Astrofísica de Andalucía (SEV-2017-0709); the Toray Science Foundation; the US Department of Energy (USDOE) through the Los Alamos National Laboratory (operated by Triad National Security, LLC, for the National Nuclear Security Administration of the USDOE (Contract 89233218CNA000001)); the Italian Ministero dell'Istruzione Università e Ricerca through the grant Progetti Premiali 2012-iALMA (CUP C52I13000140001); ALMA North America Development Fund; the European Union's Horizon 2020 research and innovation programme under grant agreement No 730562 RadioNet; Chandra TM6-17006X. This work used the Extreme Science and Engineering Discovery Environment (XSEDE), supported by NSF grant ACI-1548562, and CyVerse, supported by NSF grants DBI-0735191, DBI-1265383, and DBI-1743442. XSEDE Stampede2 resource at TACC was allocated through TG-AST170024 and TG-AST080026N. XSEDE JetStream resource at PTI and TACC was allocated through AST170028. The simulations were performed in part on the SuperMUC cluster at the LRZ in Garching, on the LOEWE cluster in CSC in Frankfurt, and on the HazelHen cluster at the HLRS in Stuttgart. This research was enabled in part by support provided by Compute Ontario (http://computeontario.ca), Calcul Quebec (http://www.calculquebec.ca) and Compute Canada (http://www.computecanada.ca). We wish to thank Eric W. Greisen for extending the capabilities of the AIPS software package, which made it possible to handle the wide-bandwidth EHT data. We acknowledge George Moellenbrock for his invaluable contribution to the CASA VLBI upgrade. We thank the staff at the participating observatories, correlation centers, and institutions for their enthusiastic support. This Letter makes use of the following ALMA data: ADS/JAO.ALMA#2016.1.01154.V. ALMA is a partnership of the European Southern Observatory (ESO; Europe, representing its member states), NSF, and National Institutes of Natural Sciences of Japan, together with National Research Council (Canada), Ministry of Science and Technology (MOST; Taiwan), Academia Sinica Institute of Astronomy and Astrophysics (ASIAA; Taiwan), and Korea Astronomy and Space Science Institute (KASI; Republic of Korea), in cooperation with the Republic of Chile. The Joint ALMA Observatory is operated by ESO, Associated Universities, Inc. (AUI)/NRAO, and the National Astronomical Observatory of Japan (NAOJ). The NRAO is a facility of the NSF operated under cooperative agreement by AUI. APEX is a collaboration between the Max-Planck-Institut für Radioastronomie (Germany), ESO, and the Onsala Space Observatory (Sweden). The SMA is a joint project between the SAO and ASIAA and is funded by the Smithsonian Institution and the Academia Sinica. The JCMT is operated by the East Asian Observatory on behalf of the NAOJ, ASIAA, and KASI, as well as the Ministry of Finance of China, Chinese Academy of Sciences, and the National Key R&D Program (No. 2017YFA0402700) of China. Additional funding support for the JCMT is provided by the Science and Technologies Facility Council (UK) and participating universities in the UK and Canada. The LMT project is a joint effort of the Instituto Nacional de Astrófisica, Óptica, y Electrónica (Mexico) and the University of Massachusetts at Amherst (USA). The IRAM 30-m telescope on Pico Veleta, Spain is operated by IRAM and supported by CNRS (Centre National de la Recherche Scientifique, France), MPG (Max-Planck-Gesellschaft, Germany) and IGN (Instituto Geográfico Nacional, Spain). The SMT is operated by the Arizona Radio Observatory, a part of the Steward Observatory of the University of Arizona, with financial support of operations from the State of Arizona and financial support for instrumentation development from the NSF. Partial SPT support is provided by the NSF Physics Frontier Center award (PHY-0114422) to the Kavli Institute of Cosmological Physics at the University of Chicago (USA), the Kavli Foundation, and the GBMF (GBMF-947). The SPT hydrogen maser was provided on loan from the GLT, courtesy of ASIAA. The SPT is supported by the National Science Foundation through grant PLR-1248097. Partial support is also provided by the NSF Physics Frontier Center grant PHY-1125897 to the Kavli Institute of Cosmological Physics at the University of Chicago, the Kavli Foundation and the Gordon and Betty Moore Foundation grant GBMF 947. The EHTC has received generous donations of FPGA chips from Xilinx Inc., under the Xilinx University Program. The EHTC has benefited from technology shared under open-source license by the Collaboration for Astronomy Signal Processing and Electronics Research (CASPER). The EHT project is grateful to T4Science and Microsemi for their assistance with Hydrogen Masers. This research has made use of NASA's Astrophysics Data System. We gratefully acknowledge the support provided by the extended staff of the ALMA, both from the inception of the ALMA Phasing Project through the observational campaigns of 2017 and 2018. We would like to thank A. Deller and W. Brisken for EHT-specific support with the use of DiFX. We acknowledge the significance that Maunakea, where the SMA and JCMT EHT stations are located, has for the indigenous Hawaiian people.

*Facilities:* EHT, ALMA, APEX, IRAM:30 m, JCMT, LMT, SMA, ARO:SMT, SPT.

*Software:* DiFX (Deller et al. 2011), CALC, PolConvert (Martí-Vidal et al. 2016), HOPS (Whitney et al. 2004), CASA (McMullin et al. 2007), AIPS (Greisen 2003), ParselTongue (Kettenis et al. 2006), GNU Parallel (Tange 2011), GILDAS, eht-imaging (Chael et al. 2016, 2018), Numpy (van der Walt et al. 2011), Scipy (Jones et al. 2001), Pandas (McKinney 2010), Astropy (The Astropy Collaboration et al. 2013, 2018), Jupyter (Kluyver et al. 2016), Matplotlib (Hunter 2007).

## Appendix
## Site and Data Issues

### A.1. Issues Requiring Mitigation

The JCMT and SMA are located within hundreds of meters of each other on Maunakea. The small natural fringe rate is insufficient to wash out unwanted signals on the JCMT–SMA baselines (to phased and single-dish SMA). The JCMT and the SMA used identical frequency setups in 2017, resulting in two types of spurious correlations. For correlations between JCMT and the SMA single-dish reference antenna (not used directly for science analysis), two narrowband terrestrial signals required special handling: one from the 1024 MHz spur tone of the R2DBEs, and a second one from the YIG oscillator tone (which is part of the LO chain) locally generated at the SMA. These signals were mitigated by flagging the affected frequency channels in post-processing.

Broadband celestial signals in the lower sideband with respect to the 220.1 GHz first LO used at the JCMT and SMA also contaminated the signal in the upper-sideband data. The differential fringe rate between upper and lower sidebands is of $O$(Hz); thus, the lower-sideband contamination averages out to zero over sufficiently long integration times. The contamination only affects the reference antenna contribution to the phased array, as other antennas are subject to 90°/270° Walsh switching (Thompson et al. 2017, Section 7.5) that removes on average the lower sideband signal over a Walsh cycle of 0.65 s. Correlations between the JCMT and SMA single-dish reference antenna thus get the full lower sideband contribution, but correlations between JCMT and SMA phased array only get $1/N$ contribution, where $N$ is the number of telescopes being phased. To avoid phase steering toward this spurious ~17% contribution to the signal, neither the SMA nor the JCMT is ever used as the reference station during atmospheric phase calibration. For scans with very small fringe-rates, there may be a small residual contribution after the 10 s averages used for network calibration (Section 6.2). This adds to the intra-site baseline amplitude error budget that propagates into gain solutions for that procedure, as well as for closure amplitudes that use the baseline on comparable timescales.

Data from PV were subject to substantial amplitude loss due to instabilities in the signal chain, attributed to excess phase noise in the maser frequency reference (which has since been replaced). Examination of the data on the ALMA–PV baseline with progressively shorter APs demonstrated a pattern of frequency spikes off the main signal with evidence that the full correlated amplitude could be recovered with an AP of 2.048 ms. Further examination of a variety of scans showed that the pattern of frequency spikes was stable across scans, sources, and days, and the amplitude loss was constant. The effect was mitigated by continuing to use the data with a 0.4 s AP and multiplying the visibility amplitudes on baselines to PV by a constant derived multiplicative factor of 1.914 during a priori flux density calibration, which is equivalent to multiplying the effective SEFD for PV by 3.663.

Misconfigured Mark 6 recorders at APEX caused substantial data loss on many scans. The first 20–30 s of recording on a particular scan (sometimes much longer) were generally good, but partial or complete data dropouts could occur thereafter. `DiFX` accounts for the amount of valid data and automatically corrects averaged amplitudes and data weights for partial data loss to within ∼1% accuracy. The remaining data from long-duration dropouts were manually flagged to avoid introducing bad APEX data into the processed data. The consequence is that ALMA–APEX coverage is inconsistent, and this complicates the strategy for network calibration and closure amplitude analysis, which makes use of intra-site baseline coverage. It also means that for the 2017 observations, APEX cannot be consistently used to help calibrate ALMA amplitude variation during poor weather when ALMA phasing efficiency is unstable.

A separate unrelated small correction factor is applied to APEX baselines to account for reduction in amplitude from the introduction of a 1 pulse-per-second (PPS) signal in the APEX data. The factor is estimated by measuring amplitudes with and without the PPS signal flagged. It is valid for multi-second averages of visibility amplitudes.

Isolated groups of frequency channels in the beamformer system at the SMA were occasionally corrupted, causing a small fraction of the bandwidth (in the high band) to be lost during the first three days of the observation. Processing of a single band within the SMA beamformer is divided across eight hardware units, each of which processes one-eighth of the total bandwidth, distributed across 128 channels of 2.234375 MHz each (Primiani et al. 2016), so that the exact pattern of lost channels, once identified, is predictable. The times when the data corruption occurred and the amount of bandwidth affected were identified using the strong noise correlation signal between the SMA (beamformed) phased array and the SMA single-dish reference (recorded on a standard EHT backend). The pattern of lost bandwidth is evenly distributed throughout the band, and we derive SEFD corrections to account for the effective relative signal power lost upon frequency average (Table 2).

The LMT data are contaminated by polarization leakage, which is delayed from the primary signal by ∼1.5 ns. This occurs in both polarizations, and is attributed to reflections in the optical setup of the LMT receiver used in 2017 (1.5 ns corresponds to 45 cm). The level of polarization leakage is ∼10%, but for an unpolarized source it will dominate the correlated signal power of cross-hand VLBI products, therefore causing a false fringe at the delayed location. During fringe closure with the HOPS pipeline, an additional 1.5 ns delay systematic is added in quadrature to LMT baselines, so that any such false fringes will not bias the global station delays. A future polarization leakage correction will need to accommodate leakage at non-zero delay to properly account for the contamination. For 2018 and beyond, the special-purpose interim receiver used at LMT was replaced by a dual-polarization sideband-separating 1.3 mm receiver with better stability and full 64 Gbps coverage with the rest of the EHT (Paper II).

### A.2. Issues not Addressed during Processing

The failure of a hard drive in one of the JCMT modules caused one-sixteenth of the data in the low band to be lost. The lost data affects all scans on the module approximately equally, as packets are scattered onto all hard drives at record time. This issue required no special handling because `DiFX` automatically adjusts data weights based on the amount of data in each AP.

Due to a small glitch in the ALMA correlator, the correlation coefficients on ALMA baselines are observed to undergo a slight dip every 18.192 s. The effective amplitude loss on scan-averaged quantities, less than 0.1%, is well within the error budget and therefore unmitigated.

No corrections were made for losses due to finite fast Fourier transform (FFT) lengths, which are required to be long in order to align ALMA $32 \times 58.59375$ MHz data in the frequency domain with the wideband 2048 MHz single-channel data from most EHT stations. A small loss is introduced due to the changing delay over the 64 μs of time corresponding to the FFT length used. The loss is zero at the DC edge of the channel and increases linearly with frequency. This effect is baseline-dependent and greatest on the baselines with the greatest east–west extent, especially when the source is rising at one location and setting at the other. Across all fringes on all sources on all baselines on all five days, the median signal loss is 0.67%, with the worst case (on a scan on the Hawai'i–PV baseline) about an order of magnitude larger. FFT losses are negligible on baselines to ALMA because the delay error accumulates over a maximum of 58.59375 MHz in frequency rather than 2048 MHz.

The LMT faces significant challenges in maintaining an accurate surface for 1.3 mm as the temperature fluctuates over the course of the evening. Pointing was also a challenge for scans at low or high elevation. These issues result in large residual gain trends obtained via amplitude self-calibration beyond the nominal error budget (Paper IV). However, the station-based amplitude gain issues do not influence robust interferometric closure quantities.

The SPT, participating for the first time in the VLBI observations, suffered from pointing problems early in the campaign. 3C 279 observing time was used to diagnose and resolve these issues, resulting in missing a majority of 3C 279 scans on April 5 and 6. The pointing issues were known and captured in observing logs during the run. The non-detections do not appear in the 3C 279 data set (Figure 2), and their absence is expected.

### A.3. Issues at Correlation

Two unanticipated issues with the ALMA data were discovered and fixed in a seventh revision (Rev7) correlation. First, the tuning of one of the ALMA LO generators was specified to insufficient precision, resulting in an undocumented 50 mHz LO offset. In most VLBI experiments, such a small LO offset might be transparently compensated by a small change in fitted delay-rate. However for the wide EHT bandwidths, the inability for a single delay-rate to model the effect over the entire 2 GHz band is noticed, where the result of imperfect correction is to imprint a small rate slope with frequency, or, equivalently, a small delay drift with time. For this reason, the effect is separately corrected for prior to fringe fitting when post-processing Rev5 data, which is possible for sufficiently small LO offsets.

Second, it was discovered that the ALMA delay system automatically removes the bulk atmospheric delay from above the array. By default, `DiFX` tries to remove the bulk atmospheric delay from above each station, resulting in a double correction for ALMA. This was most noticeable at low elevation, where the double correction imprinted a large and rapidly (but monotonically) changing delay-rate. The large residual delay-rate is not large enough to cause decoherence over the duration of a correlation AP (0.4 s). The changing delay-rate causes substantial decoherence over a several-minute scan if only a first-order fringe solution is used. Because EHT data reduction already includes a mechanism to measure and correct for nonlinear phase due to atmospheric turbulence, it can also compensate for this drift in delay-rate imprinted on the data in the initial correlation. So long as signal-to-noise is sufficient to measure phase over short timescales, the impact on calibrated data is negligible.

Both of these issues were ultimately corrected in a final Rev7 correlation release. This included the LO adjustment for ALMA as well as special scripting for the geometric model preparation that allows the normal atmospheric correction at all sites other than ALMA to be merged with a no-atmospheric correction at ALMA. Comparison of SR1 results with comparable processing of Rev7 shows no significant difference, showing that the effects were sufficiently mitigated in post-processing for SR1.

## ORCID iDs

Kazunori Akiyama https://orcid.org/0000-0002-9475-4254
Antxon Alberdi https://orcid.org/0000-0002-9371-1033
Rebecca Azulay https://orcid.org/0000-0002-2200-5393
Anne-Kathrin Baczko https://orcid.org/0000-0003-3090-3975
Mislav Baloković https://orcid.org/0000-0003-0476-6647
John Barrett https://orcid.org/0000-0002-9290-0764
Lindy Blackburn https://orcid.org/0000-0002-9030-642X
Katherine L. Bouman https://orcid.org/0000-0003-0077-4367
Geoffrey C. Bower https://orcid.org/0000-0003-4056-9982
Christiaan D. Brinkerink https://orcid.org/0000-0002-2322-0749
Roger Brissenden https://orcid.org/0000-0002-2556-0894
Silke Britzen https://orcid.org/0000-0001-9240-6734
Avery E. Broderick https://orcid.org/0000-0002-3351-760X
Do-Young Byun https://orcid.org/0000-0003-1157-4109
Andrew Chael https://orcid.org/0000-0003-2966-6220
Chi-kwan Chan https://orcid.org/0000-0001-6337-6126
Shami Chatterjee https://orcid.org/0000-0002-2878-1502
Ilje Cho https://orcid.org/0000-0001-6083-7521
Pierre Christian https://orcid.org/0000-0001-6820-9941
John E. Conway https://orcid.org/0000-0003-2448-9181
Geoffrey B. Crew https://orcid.org/0000-0002-2079-3189
Yuzhu Cui https://orcid.org/0000-0001-6311-4345
Jordy Davelaar https://orcid.org/0000-0002-2685-2434
Mariafelicia De Laurentis https://orcid.org/0000-0002-9945-682X
Roger Deane https://orcid.org/0000-0003-1027-5043
Jessica Dempsey https://orcid.org/0000-0003-1269-9667
Gregory Desvignes https://orcid.org/0000-0003-3922-4055
Jason Dexter https://orcid.org/0000-0003-3903-0373
Sheperd S. Doeleman https://orcid.org/0000-0002-9031-0904
Ralph P. Eatough https://orcid.org/0000-0001-6196-4135
Heino Falcke https://orcid.org/0000-0002-2526-6724
Vincent L. Fish https://orcid.org/0000-0002-7128-9345
Raquel Fraga-Encinas https://orcid.org/0000-0002-5222-1361
José L. Gómez https://orcid.org/0000-0003-4190-7613
Peter Galison https://orcid.org/0000-0002-6429-3872
Charles F. Gammie https://orcid.org/0000-0001-7451-8935
Boris Georgiev https://orcid.org/0000-0002-3586-6424
Roman Gold https://orcid.org/0000-0003-2492-1966
Minfeng Gu (顾敏峰) https://orcid.org/0000-0002-4455-6946
Mark Gurwell https://orcid.org/0000-0003-0685-3621
Kazuhiro Hada https://orcid.org/0000-0001-6906-772X
Ronald Hesper https://orcid.org/0000-0003-1918-6098
Luis C. Ho (何子山) https://orcid.org/0000-0001-6947-5846
Mareki Honma https://orcid.org/0000-0003-4058-9000
Chih-Wei L. Huang https://orcid.org/0000-0001-5641-3953
Shiro Ikeda https://orcid.org/0000-0002-2462-1448
Sara Issaoun https://orcid.org/0000-0002-5297-921X
David J. James https://orcid.org/0000-0001-5160-4486
Michael Janssen https://orcid.org/0000-0001-8685-6544
Britton Jeter https://orcid.org/0000-0003-2847-1712
Wu Jiang (江悟) https://orcid.org/0000-0001-7369-3539
Michael D. Johnson https://orcid.org/0000-0002-4120-3029
Svetlana Jorstad https://orcid.org/0000-0001-6158-1708
Taehyun Jung https://orcid.org/0000-0001-7003-8643
Mansour Karami https://orcid.org/0000-0001-7387-9333
Ramesh Karuppusamy https://orcid.org/0000-0002-5307-2919
Tomohisa Kawashima https://orcid.org/0000-0001-8527-0496
Garrett K. Keating https://orcid.org/0000-0002-3490-146X
Mark Kettenis https://orcid.org/0000-0002-6156-5617
Jae-Young Kim https://orcid.org/0000-0001-8229-7183
Junhan Kim https://orcid.org/0000-0002-4274-9373
Motoki Kino https://orcid.org/0000-0002-2709-7338
Jun Yi Koay https://orcid.org/0000-0002-7029-6658
Patrick M. Koch https://orcid.org/0000-0003-2777-5861
Shoko Koyama https://orcid.org/0000-0002-3723-3372
Michael Kramer https://orcid.org/0000-0002-4175-2271
Carsten Kramer https://orcid.org/0000-0002-4908-4925
Thomas P. Krichbaum https://orcid.org/0000-0002-4892-9586
Tod R. Lauer https://orcid.org/0000-0003-3234-7247
Sang-Sung Lee https://orcid.org/0000-0002-6269-594X
Yan-Rong Li (李彦荣) https://orcid.org/0000-0001-5841-9179
Zhiyuan Li (李志远) https://orcid.org/0000-0003-0355-6437
Michael Lindqvist https://orcid.org/0000-0002-3669-0715
Kuo Liu https://orcid.org/0000-0002-2953-7376
Elisabetta Liuzzo https://orcid.org/0000-0003-0995-5201
Laurent Loinard https://orcid.org/0000-0002-5635-3345
Ru-Sen Lu (路如森) https://orcid.org/0000-0002-7692-7967
Nicholas R. MacDonald https://orcid.org/0000-0002-6684-8691
Jirong Mao (毛基荣) https://orcid.org/0000-0002-7077-7195
Sera Markoff https://orcid.org/0000-0001-9564-0876
Daniel P. Marrone https://orcid.org/0000-0002-2367-1080
Alan P. Marscher https://orcid.org/0000-0001-7396-3332
Iván Martí-Vidal https://orcid.org/0000-0003-3708-9611
Lynn D. Matthews https://orcid.org/0000-0002-3728-8082
Lia Medeiros https://orcid.org/0000-0003-2342-6728
Karl M. Menten https://orcid.org/0000-0001-6459-0669
Yosuke Mizuno https://orcid.org/0000-0002-8131-6730
Izumi Mizuno https://orcid.org/0000-0002-7210-6264
James M. Moran https://orcid.org/0000-0002-3882-4414
Kotaro Moriyama https://orcid.org/0000-0003-1364-3761
Monika Moscibrodzka https://orcid.org/0000-0002-4661-6332
Cornelia Müller https://orcid.org/0000-0002-2739-2994
Hiroshi Nagai https://orcid.org/0000-0003-0292-3645
Neil M. Nagar https://orcid.org/0000-0001-6920-662X
Masanori Nakamura https://orcid.org/0000-0001-6081-2420

Ramesh Narayan  https://orcid.org/0000-0002-1919-2730
Iniyan Natarajan  https://orcid.org/0000-0001-8242-4373
Chunchong Ni  https://orcid.org/0000-0003-1361-5699
Aristeidis Noutsos  https://orcid.org/0000-0002-4151-3860
Héctor Olivares  https://orcid.org/0000-0001-6833-7580
Gisela N. Ortiz-León  https://orcid.org/0000-0002-2863-676X
Daniel C. M. Palumbo  https://orcid.org/0000-0002-7179-3816
Ue-Li Pen  https://orcid.org/0000-0003-2155-9578
Dominic W. Pesce  https://orcid.org/0000-0002-5278-9221
Oliver Porth  https://orcid.org/0000-0002-4584-2557
Ben Prather  https://orcid.org/0000-0002-0393-7734
Jorge A. Preciado-López  https://orcid.org/0000-0002-4146-0113
Hung-Yi Pu  https://orcid.org/0000-0001-9270-8812
Venkatessh Ramakrishnan  https://orcid.org/0000-0002-9248-086X
Ramprasad Rao  https://orcid.org/0000-0002-1407-7944
Alexander W. Raymond  https://orcid.org/0000-0002-5779-4767
Luciano Rezzolla  https://orcid.org/0000-0002-1330-7103
Bart Ripperda  https://orcid.org/0000-0002-7301-3908
Freek Roelofs  https://orcid.org/0000-0001-5461-3687
Eduardo Ros  https://orcid.org/0000-0001-9503-4892
Mel Rose  https://orcid.org/0000-0002-2016-8746
Alan L. Roy  https://orcid.org/0000-0002-1931-0135
Chet Ruszczyk  https://orcid.org/0000-0001-7278-9707
Benjamin R. Ryan  https://orcid.org/0000-0001-8939-4461
Kazi L. J. Rygl  https://orcid.org/0000-0003-4146-9043
David Sánchez-Arguelles  https://orcid.org/0000-0002-7344-9920
Mahito Sasada  https://orcid.org/0000-0001-5946-9960
Tuomas Savolainen  https://orcid.org/0000-0001-6214-1085
Lijing Shao  https://orcid.org/0000-0002-1334-8853
Zhiqiang Shen (沈志强)  https://orcid.org/0000-0003-3540-8746
Des Small  https://orcid.org/0000-0003-3723-5404
Bong Won Sohn  https://orcid.org/0000-0002-4148-8378
Jason SooHoo  https://orcid.org/0000-0003-1938-0720
Fumie Tazaki  https://orcid.org/0000-0003-0236-0600
Paul Tiede  https://orcid.org/0000-0003-3826-5648
Remo P. J. Tilanus  https://orcid.org/0000-0002-6514-553X
Michael Titus  https://orcid.org/0000-0002-3423-4505
Kenji Toma  https://orcid.org/0000-0002-7114-6010
Pablo Torne  https://orcid.org/0000-0001-8700-6058
Sascha Trippe  https://orcid.org/0000-0003-0465-1559
Ilse van Bemmel  https://orcid.org/0000-0001-5473-2950
Huib Jan van Langevelde  https://orcid.org/0000-0002-0230-5946
Daniel R. van Rossum  https://orcid.org/0000-0001-7772-6131
John Wardle  https://orcid.org/0000-0002-8960-2942
Jonathan Weintroub  https://orcid.org/0000-0002-4603-5204
Norbert Wex  https://orcid.org/0000-0003-4058-2837
Robert Wharton  https://orcid.org/0000-0002-7416-5209
Maciek Wielgus  https://orcid.org/0000-0002-8635-4242
George N. Wong  https://orcid.org/0000-0001-6952-2147
Qingwen Wu (吴庆文)  https://orcid.org/0000-0003-4773-4987
André Young  https://orcid.org/0000-0003-0000-2682
Ken Young  https://orcid.org/0000-0002-3666-4920
Ziri Younsi  https://orcid.org/0000-0001-9283-1191
Feng Yuan (袁峰)  https://orcid.org/0000-0003-3564-6437
J. Anton Zensus  https://orcid.org/0000-0001-7470-3321
Guangyao Zhao  https://orcid.org/0000-0002-4417-1659

Shan-Shan Zhao  https://orcid.org/0000-0002-9774-3606
Joseph R. Farah  https://orcid.org/0000-0003-4914-5625
Daniel Michalik  https://orcid.org/0000-0002-7618-6556
Andrew Nadolski  https://orcid.org/0000-0001-9479-9957
Rurik A. Primiani  https://orcid.org/0000-0003-3910-7529
Paul Yamaguchi  https://orcid.org/0000-0002-6017-8199

---

The Event Horizon Telescope Collaboration,

Kazunori Akiyama[1,2,3,4] , Antxon Alberdi[5] , Walter Alef[6], Keiichi Asada[7], Rebecca Azulay[8,9,6] , Anne-Kathrin Baczko[6] , David Ball[10], Mislav Baloković[4,11] , John Barrett[2] , Dan Bintley[12], Lindy Blackburn[4,11] , Wilfred Boland[13], Katherine L. Bouman[4,11,14] , Geoffrey C. Bower[15] , Michael Bremer[16], Christiaan D. Brinkerink[17] , Roger Brissenden[4,11] , Silke Britzen[6] , Avery E. Broderick[18,19,20] , Dominique Broguiere[16], Thomas Bronzwaer[17], Do-Young Byun[21,22] , John E. Carlstrom[23,24,25,26], Andrew Chael[4,11] , Chi-kwan Chan[10,27] , Shami Chatterjee[28] , Koushik Chatterjee[29], Ming-Tang Chen[15], Yongjun Chen (陈永军)[30,31], Ilje Cho[21,22] , Pierre Christian[10,11], John E. Conway[32] , James M. Cordes[28], Geoffrey B. Crew[2] , Yuzhu Cui[33,34] , Jordy Davelaar[17] , Mariafelicia De Laurentis[35,36,37] , Roger Deane[38,39] , Jessica Dempsey[12] , Gregory Desvignes[6] , Jason Dexter[40] , Sheperd S. Doeleman[4,11] , Ralph P. Eatough[6] , Heino Falcke[17] , Vincent L. Fish[2] , Ed Fomalont[1], Raquel Fraga-Encinas[17] , Per Friberg[12], Christian M. Fromm[36], José L. Gómez[5] , Peter Galison[4,41,42] , Charles F. Gammie[43,44] , Roberto García[16], Olivier Gentaz[16], Boris Georgiev[19] , Ciriaco Goddi[17,45], Roman Gold[36] , Minfeng Gu (顾敏峰)[30,46] , Mark Gurwell[11] , Kazuhiro Hada[33,34] , Michael H. Hecht[2], Ronald Hesper[47] , Luis C. Ho (何子山)[48,49] , Paul Ho[7], Mareki Honma[33,34] , Chih-Wei L. Huang[7] , Lei Huang (黄磊)[30,46] , David H. Hughes[50], Shiro Ikeda[3,51,52,53] , Makoto Inoue[7], Sara Issaoun[17] , David J. James[4,11] , Buell T. Jannuzi[10], Michael Janssen[17] , Britton Jeter[19,20] , Wu Jiang (江悟)[30] , Michael D. Johnson[4,11] , Svetlana Jorstad[54,55] , Taehyun Jung[21,22] , Mansour Karami[18,19] , Ramesh Karuppusamy[6] , Tomohisa Kawashima[3] , Garrett K. Keating[11] , Mark Kettenis[56] , Jae-Young Kim[6] , Junhan Kim[10] , Jongsoo Kim[21], Motoki Kino[3,57] , Jun Yi Koay[7] , Patrick M. Koch[7] , Shoko Koyama[7] , Michael Kramer[6] , Carsten Kramer[16] , Thomas P. Krichbaum[6] , Cheng-Yu Kuo[58], Tod R. Lauer[59] , Sang-Sung Lee[21] , Yan-Rong Li (李彦荣)[60] , Zhiyuan Li (李志远)[61,62] , Michael Lindqvist[32] , Kuo Liu[6] , Elisabetta Liuzzo[63] , Wen-Ping Lo[7,64], Andrei P. Lobanov[6], Laurent Loinard[65,66] , Colin Lonsdale[2], Ru-Sen Lu (路如森)[30,6] , Nicholas R. MacDonald[6] , Jirong Mao (毛基荣)[67,68,69] , Sera Markoff[29,70] , Daniel P. Marrone[10] , Alan P. Marscher[54] , Iván Martí-Vidal[32,71] , Satoki Matsushita[7], Lynn D. Matthews[2] , Lia Medeiros[10,72] , Karl M. Menten[6], Yosuke Mizuno[36] , Izumi Mizuno[12] , James M. Moran[4,11] , Kotaro Moriyama[33,2] , Monika Moscibrodzka[17] , Cornelia Müller[6,17] , Hiroshi Nagai[3,34] , Neil M. Nagar[73] , Masanori Nakamura[7] , Ramesh Narayan[4,11] , Gopal Narayanan[74], Iniyan Natarajan[39] , Roberto Neri[16] , Chunchong Ni[19,20] , Aristeidis Noutsos[6] , Hiroki Okino[33,75], Héctor Olivares[36] , Gisela N. Ortiz-León[6] , Tomoaki Oyama[33], Feryal Özel[10], Daniel C. M. Palumbo[4,11] , Nimesh Patel[11], Ue-Li Pen[18,76,77,78] , Dominic W. Pesce[4,11] , Vincent Piétu[16], Richard Plambeck[79], Aleksandar PopStefanija[74], Oliver Porth[36,29] , Ben Prather[43], Jorge A. Preciado-López[18] , Dimitrios Psaltis[10], Hung-Yi Pu[18] , Venkatessh Ramakrishnan[73] , Ramprasad Rao[15] , Mark G. Rawlings[12], Alexander W. Raymond[4,11] , Luciano Rezzolla[36] , Bart Ripperda[36] , Freek Roelofs[17] , Alan Rogers[2], Eduardo Ros[6] , Mel Rose[10] , Arash Roshanineshat[10], Helge Rottmann[6], Alan L. Roy[6] , Chet Ruszczyk[2] , Benjamin R. Ryan[80,81] , Kazi L. J. Rygl[63] , Salvador Sánchez[82], David Sánchez-Arguelles[50,83] , Mahito Sasada[33,84] , Tuomas Savolainen[6,85,86] , F. Peter Schloerb[74], Karl-Friedrich Schuster[16], Lijing Shao[6,49] , Zhiqiang Shen (沈志强)[30,31] , Des Small[56] , Bong Won Sohn[21,22,87] , Jason SooHoo[2] , Fumie Tazaki[33] , Paul Tiede[18,19] , Remo P. J. Tilanus[17,45,88] , Michael Titus[2] , Kenji Toma[89,90] , Pablo Torne[6,82] , Tyler Trent[10], Sascha Trippe[91] , Shuichiro Tsuda[33], Ilse van Bemmel[56] , Huib Jan van Langevelde[56,92] , Daniel R. van Rossum[17] ,

Jan Wagner[6], John Wardle[93] , Jonathan Weintraub[4,11], Norbert Wex[6] , Robert Wharton[6] , Maciek Wielgus[4,11] , George N. Wong[43] , Qingwen Wu (吴庆文)[94] , André Young[17] , Ken Young[11], Ziri Younsi[95,36] , Feng Yuan (袁峰)[30,46,96], Ye-Fei Yuan (袁业飞)[97], J. Anton Zensus[6] , Guangyao Zhao[21] , Shan-Shan Zhao[17,61] , Ziyan Zhu[42], Roger Cappallo[2], Joseph R. Farah[11,98,4] , Thomas W. Folkers[10], Zheng Meyer-Zhao[7,99] , Daniel Michalik[100,101] , Andrew Nadolski[44] , Hiroaki Nishioka[7], Nicolas Pradel[7], Rurik A. Primiani[11,102] , Kamal Souccar[74], Laura Vertatschitsch[11,102], and Paul Yamaguchi[11]

1 National Radio Astronomy Observatory, 520 Edgemont Rd, Charlottesville, VA 22903, USA
2 Massachusetts Institute of Technology Haystack Observatory, 99 Millstone Road, Westford, MA 01886, USA
3 National Astronomical Observatory of Japan, 2-21-1 Osawa, Mitaka, Tokyo 181-8588, Japan
4 Black Hole Initiative at Harvard University, 20 Garden Street, Cambridge, MA 02138, USA
5 Instituto de Astrofísica de Andalucía-CSIC, Glorieta de la Astronomía s/n, E-18008 Granada, Spain
6 Max-Planck-Institut für Radioastronomie, Auf dem Hügel 69, D-53121 Bonn, Germany
7 Institute of Astronomy and Astrophysics, Academia Sinica, 11F of Astronomy-Mathematics Building,
AS/NTU No. 1, Sec. 4, Roosevelt Rd, Taipei 10617, Taiwan, R.O.C.
8 Departament d'Astronomia i Astrofísica, Universitat de València, C. Dr. Moliner 50, E-46100 Burjassot, València, Spain
9 Observatori Astronòmic, Universitat de València, C. Catedrático José Beltrán 2, E-46980 Paterna, València, Spain
10 Steward Observatory and Department of Astronomy, University of Arizona, 933 N. Cherry Avenue, Tucson, AZ 85721, USA
11 Center for Astrophysics | Harvard & Smithsonian, 60 Garden Street, Cambridge, MA 02138, USA
12 East Asian Observatory, 660 N. A'ohoku Pl., Hilo, HI 96720, USA
13 Nederlandse Onderzoekschool voor Astronomie (NOVA), PO Box 9513, 2300 RA Leiden, The Netherlands
14 California Institute of Technology, 1200 East California Boulevard, Pasadena, CA 91125, USA
15 Institute of Astronomy and Astrophysics, Academia Sinica, 645 N. A'ohoku Place, Hilo, HI 96720, USA
16 Institut de Radioastronomie Millimétrique, 300 rue de la Piscine, F-38406 Saint Martin d'Hères, France
17 Department of Astrophysics, Institute for Mathematics, Astrophysics and Particle Physics (IMAPP),
Radboud University, P.O. Box 9010, 6500 GL Nijmegen, The Netherlands
18 Perimeter Institute for Theoretical Physics, 31 Caroline Street North, Waterloo, ON, N2L 2Y5, Canada
19 Department of Physics and Astronomy, University of Waterloo, 200 University Avenue West, Waterloo, ON, N2L 3G1, Canada
20 Waterloo Centre for Astrophysics, University of Waterloo, Waterloo, ON N2L 3G1, Canada
21 Korea Astronomy and Space Science Institute, Daedeok-daero 776, Yuseong-gu, Daejeon 34055, Republic of Korea
22 University of Science and Technology, Gajeong-ro 217, Yuseong-gu, Daejeon 34113, Republic of Korea
23 Kavli Institute for Cosmological Physics, University of Chicago, 5640 South Ellis Avenue, Chicago, IL 60637, USA
24 Department of Astronomy and Astrophysics, University of Chicago, 5640 South Ellis Avenue, Chicago, IL 60637, USA
25 Department of Physics, University of Chicago, 5720 South Ellis Avenue, Chicago, IL 60637, USA
26 Enrico Fermi Institute, University of Chicago, 5640 South Ellis Avenue, Chicago, IL 60637, USA
27 Data Science Institute, University of Arizona, 1230 N. Cherry Ave., Tucson, AZ 85721, USA
28 Cornell Center for Astrophysics and Planetary Science, Cornell University, Ithaca, NY 14853, USA
29 Anton Pannekoek Institute for Astronomy, University of Amsterdam, Science Park 904, 1098 XH, Amsterdam, The Netherlands
30 Shanghai Astronomical Observatory, Chinese Academy of Sciences, 80 Nandan Road, Shanghai 200030, People's Republic of China
31 Key Laboratory of Radio Astronomy, Chinese Academy of Sciences, Nanjing 210008, People's Republic of China
32 Department of Space, Earth and Environment, Chalmers University of Technology, Onsala Space Observatory, SE-43992 Onsala, Sweden
33 Mizusawa VLBI Observatory, National Astronomical Observatory of Japan, 2-12 Hoshigaoka, Mizusawa, Oshu, Iwate 023-0861, Japan
34 Department of Astronomical Science, The Graduate University for Advanced Studies (SOKENDAI), 2-21-1 Osawa, Mitaka, Tokyo 181-8588, Japan
35 Dipartimento di Fisica "E. Pancini", Universitá di Napoli "Federico II", Compl. Univ. di Monte S. Angelo, Edificio G, Via Cinthia, I-80126, Napoli, Italy
36 Institut für Theoretische Physik, Goethe-Universität Frankfurt, Max-von-Laue-Straße 1, D-60438 Frankfurt am Main, Germany
37 INFN Sez. di Napoli, Compl. Univ. di Monte S. Angelo, Edificio G, Via Cinthia, I-80126, Napoli, Italy
38 Department of Physics, University of Pretoria, Lynnwood Road, Hatfield, Pretoria 0083, South Africa
39 Centre for Radio Astronomy Techniques and Technologies, Department of Physics and Electronics, Rhodes University, Grahamstown 6140, South Africa
40 Max-Planck-Institut für Extraterrestrische Physik, Giessenbachstr. 1, D-85748 Garching, Germany
41 Department of History of Science, Harvard University, Cambridge, MA 02138, USA
42 Department of Physics, Harvard University, Cambridge, MA 02138, USA
43 Department of Physics, University of Illinois, 1110 West Green St, Urbana, IL 61801, USA
44 Department of Astronomy, University of Illinois at Urbana-Champaign, 1002 West Green Street, Urbana, Illinois 61801, USA
45 Leiden Observatory—Allegro, Leiden University, P.O. Box 9513, 2300 RA Leiden, The Netherlands
46 Key Laboratory for Research in Galaxies and Cosmology, Chinese Academy of Sciences, Shanghai 200030, People's Republic of China
47 NOVA Sub-mm Instrumentation Group, Kapteyn Astronomical Institute, University of Groningen, Landleven 12, 9747 AD Groningen, The Netherlands
48 Department of Astronomy, School of Physics, Peking University, Beijing 100871, People's Republic of China
49 Kavli Institute for Astronomy and Astrophysics, Peking University, Beijing 100871, People's Republic of China
50 Instituto Nacional de Astrofísica, Óptica y Electrónica. Apartado Postal 51 y 216, 72000. Puebla Pue., México
51 The Institute of Statistical Mathematics, 10-3 Midori-cho, Tachikawa, Tokyo, 190-8562, Japan
52 Department of Statistical Science, The Graduate University for Advanced Studies (SOKENDAI), 10-3 Midori-cho, Tachikawa, Tokyo 190-8562, Japan
53 Kavli Institute for the Physics and Mathematics of the Universe, The University of Tokyo, 5-1-5 Kashiwanoha, Kashiwa, 277-8583, Japan
54 Institute for Astrophysical Research, Boston University, 725 Commonwealth Ave., Boston, MA 02215, USA
55 Astronomical Institute, St. Petersburg University, Universitetskij pr., 28, Petrodvorets,198504 St.Petersburg, Russia
56 Joint Institute for VLBI ERIC (JIVE), Oude Hoogeveensedijk 4, 7991 PD Dwingeloo, The Netherlands
57 Kogakuin University of Technology and Engineering, Academic Support Center, 2665-1 Nakano, Hachioji, Tokyo 192-0015, Japan
58 Physics Department, National Sun Yat-Sen University, No. 70, Lien-Hai Rd, Kaosiung City 80424, Taiwan, R.O.C
59 National Optical Astronomy Observatory, 950 North Cherry Ave., Tucson, AZ 85719, USA
60 Key Laboratory for Particle Astrophysics, Institute of High Energy Physics, Chinese Academy of Sciences,
19B Yuquan Road, Shijingshan District, Beijing, People's Republic of China
61 School of Astronomy and Space Science, Nanjing University, Nanjing 210023, People's Republic of China
62 Key Laboratory of Modern Astronomy and Astrophysics, Nanjing University, Nanjing 210023, People's Republic of China
63 Italian ALMA Regional Centre, INAF-Istituto di Radioastronomia, Via P. Gobetti 101, I-40129 Bologna, Italy
64 Department of Physics, National Taiwan University, No.1, Sect.4, Roosevelt Rd., Taipei 10617, Taiwan, R.O.C

[65] Instituto de Radioastronomía y Astrofísica, Universidad Nacional Autónoma de México, Morelia 58089, México

[66] Instituto de Astronomía, Universidad Nacional Autónoma de México, CdMx 04510, México

[67] Yunnan Observatories, Chinese Academy of Sciences, 650011 Kunming, Yunnan Province, People's Republic of China

[68] Center for Astronomical Mega-Science, Chinese Academy of Sciences, 20A Datun Road, Chaoyang District, Beijing, 100012, People's Republic of China

[69] Key Laboratory for the Structure and Evolution of Celestial Objects, Chinese Academy of Sciences, 650011 Kunming, People's Republic of China

[70] Gravitation Astroparticle Physics Amsterdam (GRAPPA) Institute, University of Amsterdam, Science Park 904, 1098 XH Amsterdam, The Netherlands

[71] Centro Astronómico de Yebes (IGN), Apartado 148, E-19180 Yebes, Spain

[72] Department of Physics, Broida Hall, University of California Santa Barbara, Santa Barbara, CA 93106, USA

[73] Astronomy Department, Universidad de Concepción, Casilla 160-C, Concepción, Chile

[74] Department of Astronomy, University of Massachusetts, 01003, Amherst, MA, USA

[75] Department of Astronomy, Graduate School of Science, The University of Tokyo, 7-3-1 Hongo, Bunkyo-ku, Tokyo 113-0033, Japan

[76] Canadian Institute for Theoretical Astrophysics, University of Toronto, 60 St. George Street, Toronto, ON M5S 3H8, Canada

[77] Dunlap Institute for Astronomy and Astrophysics, University of Toronto, 50 St. George Street, Toronto, ON M5S 3H4, Canada

[78] Canadian Institute for Advanced Research, 180 Dundas St West, Toronto, ON M5G 1Z8, Canada

[79] Radio Astronomy Laboratory, University of California, Berkeley, CA 94720, USA

[80] CCS-2, Los Alamos National Laboratory, P.O. Box 1663, Los Alamos, NM 87545, USA

[81] Center for Theoretical Astrophysics, Los Alamos National Laboratory, Los Alamos, NM, 87545, USA

[82] Instituto de Radioastronomía Milimétrica, IRAM, Avenida Divina Pastora 7, Local 20, E-18012, Granada, Spain

[83] Consejo Nacional de Ciencia y Tecnología, Av. Insurgentes Sur 1582, 03940, Ciudad de México, México

[84] Hiroshima Astrophysical Science Center, Hiroshima University, 1-3-1 Kagamiyama, Higashi-Hiroshima, Hiroshima 739-8526, Japan

[85] Aalto University Department of Electronics and Nanoengineering, PL 15500, FI-00076 Aalto, Finland

[86] Aalto University Metsähovi Radio Observatory, Metsähovintie 114, FI-02540 Kylmälä, Finland

[87] Department of Astronomy, Yonsei University, Yonsei-ro 50, Seodaemun-gu, 03722 Seoul, Republic of Korea

[88] Netherlands Organisation for Scientific Research (NWO), Postbus 93138, 2509 AC Den Haag, The Netherlands

[89] Frontier Research Institute for Interdisciplinary Sciences, Tohoku University, Sendai 980-8578, Japan

[90] Astronomical Institute, Tohoku University, Sendai 980-8578, Japan

[91] Department of Physics and Astronomy, Seoul National University, Gwanak-gu, Seoul 08826, Republic of Korea

[92] Leiden Observatory, Leiden University, Postbus 2300, 9513 RA Leiden, The Netherlands

[93] Physics Department, Brandeis University, 415 South Street, Waltham, MA 02453, USA

[94] School of Physics, Huazhong University of Science and Technology, Wuhan, Hubei, 430074, People's Republic of China

[95] Mullard Space Science Laboratory, University College London, Holmbury St. Mary, Dorking, Surrey, RH5 6NT, UK

[96] School of Astronomy and Space Sciences, University of Chinese Academy of Sciences, No. 19A Yuquan Road, Beijing 100049, People's Republic of China

[97] Astronomy Department, University of Science and Technology of China, Hefei 230026, People's Republic of China

[98] University of Massachusetts Boston, 100 William T, Morrissey Blvd, Boston, MA 02125, USA

[99] ASTRON, Oude Hoogeveensedijk 4, 7991 PD Dwingeloo, The Netherlands

[100] Science Support Office, Directorate of Science, European Space Research and Technology Centre (ESA/ESTEC), Keplerlaan 1, 2201 AZ Noordwijk, The Netherlands

[101] University of Chicago, 5640 South Ellis Avenue, Chicago, IL 60637, USA

[102] Systems and Technology Research, 600 West Cummings Park, Woburn, MA 01801, USA